\documentclass[11pt]{article}
\usepackage[utf8]{inputenc}
\usepackage[english]{babel}

\usepackage{amssymb}
\usepackage{mathtools}
\usepackage{bbm}

\usepackage{setspace}
\usepackage{fullpage}
\usepackage{graphicx}
\usepackage{subcaption}
\usepackage{xcolor}
\usepackage{pdflscape}
\usepackage{rotating}
\usepackage{placeins}

\usepackage{tikz}
\usepackage{pgfplots}
\usetikzlibrary{shapes.geometric, matrix, arrows.meta}

\usepackage{booktabs}
\usepackage{multirow}
\usepackage{siunitx}
\usepackage{enumitem}

\usepackage[backref=page,breaklinks=true,hidelinks]{hyperref}
\hypersetup{
    colorlinks,
    linkcolor={red!50!black},
    citecolor={blue!50!black},
    urlcolor={blue!80!black}
}
\renewcommand*{\backref}[1]{}
\renewcommand*{\backrefalt}[4]{%
    \ifcase #1 {\footnotesize(Not cited.)}%
    \or        {\footnotesize(Cited on page~#2.)}%
    \else      {\footnotesize(Cited on pages~#2.)}%
    \fi}

\usepackage{natbib}
\usepackage{amsthm}
\usepackage{thmtools}
\usepackage{thm-restate}

\usepackage{comment}
\usepackage{versions}
\usepackage[normalem]{ulem}
\usepackage{todonotes}
\presetkeys{todonotes}{inline}{}
\presetkeys{todonotes}{color=blue!10}{}

\declaretheorem[name=Theorem]{thm}
\declaretheorem[name=Proposition]{prop}
\declaretheorem[name=Assumption]{asm}

\declaretheorem[name=Algorithm]{alg}
\declaretheorem[name=Example]{exm}

\addto\extrasenglish{%

}

\makeatletter
\def\blfootnote{\xdef\@thefnmark{}\@footnotetext}
\makeatother

\newcommand{\E}{\mathbb{E}}
\renewcommand{\P}{\mathbb{P}}

\newcommand{\Cov}{\mathbb{C}\textnormal{ov}}

\newcommand{\Neutral}{{{\color{gray}\ensuremath{\varnothing}}}}
\newcommand{\fdecision}{f^{\textnormal{auto}}}
\newcommand{\ftriage}{f^{\textnormal{triage}}}

\newcommand{\Dnull}{\ensuremath{\underline{D}}}
\newcommand{\Dactive}{\ensuremath{A}}
\newcommand{\F}{\mathcal{F}}

\newcommand{\Ind}{\mathbf{1}}

\DeclareMathOperator*{\argmax}{arg\,max}

\graphicspath{{images/}}

\includeversion{nonblind}

\title{Designing Algorithmic Recommendations \\ to Achieve Human--AI Complementarity}
\begin{nonblind}
\author{
    Bryce McLaughlin  \\ {\small Wharton}
    \and
    Jann Spiess \\ {\small Stanford GSB}
}
\end{nonblind}

\date{
First public version: January 2024 \\
This version: November 2026}

\begin{document}

\maketitle

\begin{abstract}
    Algorithms frequently assist, rather than replace, human decision-makers.
    However, the design and analysis of algorithms often focus on predicting outcomes and do not explicitly model their effect on human decisions.
    This discrepancy between the design and role of algorithmic assistants becomes particularly concerning in light of empirical evidence that suggests that algorithmic assistants again and again fail to improve human decisions.
    In this article, we formalize the design of recommendation algorithms that assist human decision-makers without making restrictive ex-ante assumptions about how recommendations affect decisions.
    We formulate an algorithmic-design problem that leverages the potential-outcomes framework from causal inference to model the effect of recommendations on a human decision-maker's binary treatment choice.
    Within this model, we make a connection between algorithmic recommendations and instrumental variables.
    Leveraging this connection, we introduce a monotonicity assumption that leads to an intuitive classification of human responses to the algorithm.
    Under this assumption, we can express the human's response to algorithmic recommendations in terms of their compliance with the algorithm and the active decision they would take if the algorithm sends no recommendation.
    We showcase the utility of our framework using an online experiment that simulates a hiring task.
    We argue that our approach can make sense of the relative performance of different recommendation algorithms in the experiment and can help design solutions that realize human--AI complementarity.
    Finally, we leverage our approach to derive maximin-optimal recommendation algorithms that can be implemented with machine learning using limited training data.
\end{abstract}
\blfootnote{
        Bryce McLaughlin (\href{mailto:bryce.mclaughlin@bc.edu}{bryce.mclaughlin@bc.edu}), Carroll School of Management, Boston College and Jann Spiess (\href{mailto:jspiess@stanford.edu}{jspiess@stanford.edu}), Graduate School of Business, Stanford University.
        We thank Maya Balakrishnan, Haosen Ge, Yagan Hazard, Sanmi Koyejo, Carl Meyer, Adrienne Propp, Emma Rockall, Omer Shiran-Cohen, Jason Weitze, and audience members at the 2023, 2024 \& 2025 INFORMS Annual Meetings, 2024 MSOM Conference, 2025 Machine Learning in Economics Summer Institute, 2025 AOM Annual Meeting, Stanford, Wharton, Berkeley, MIT, Boston University, HBS, Columbia, and NYU for helpful comments and suggestions.
        We used AI tools (Claude Opus 5.5 and Claude Fable 5.1) in November 2026 for help with editing and the development of additional proofs. All AI edits were reviewed by the authors.
    }

\newpage

\section{Introduction}

Algorithms often assist, rather than replace, human decision-makers.
For example, in the United States, federal pretrial service officers set bail for defendants after observing a pretrial risk assessment \citep{cadigan_implementing_2011}, rather than letting the algorithm take a bail decision automatically. Similarly, doctors may be assisted by clinical decision support systems throughout the patient care process \citep{sutton_overview_2020} and managers may be assisted by pre-employment assessments throughout the hiring process \citep{raghavan_mitigating_2020}.
In all of these cases, the human maintains the final decision authority.
However, the design and analysis of these algorithms often focus on the prediction of outcomes as if algorithmic recommendations were implemented directly, rather than on how these recommendations affect human decisions.

The narrow focus of existing approaches on predicting outcomes rather than optimizing for human--AI complementarity
becomes of particular concern in light of a growing body of evidence that shows that human decision-makers with access to algorithmic recommendations often make worse decisions than if these recommendations had been implemented directly and, in some cases, may even make worse decisions than they would have taken without access to the algorithm. \cite{hemmer_human-ai_2021} identifies 53 articles that report (i) unassisted human performance, (ii) algorithm performance, and (iii) algorithm-assisted human performance on a decision task. Only in 16 of these does the algorithm-assisted human decision-maker perform best. The meta-analysis of \citet{vaccaro_when_2024} finds that on average human--AI combinations perform worse than the better of human and AI alone, with the losses concentrated in decision tasks. In many of these studies, decision-makers exhibit noisy and biased adoption of the algorithm's recommendations, which counteracts the gains associated with the additional information the algorithm provides. This suggests the necessity to incorporate decision-makers' adoption decisions explicitly when designing the algorithm.

This article formalizes the design of recommendation algorithms that assist humans in making binary treatment choices via an optimization problem that leverages the potential-outcomes framework from causal inference.
To analyze recommendation algorithms, we employ the notion of ``potential decisions'' and make connections to the analysis of instrumental variables in causal inference with imperfect compliance. This view suggests that an optimal recommendation algorithm should take into account the final outcome of interest to which the decision is related \emph{as well as} the decision-maker's choice of when to comply with a recommendation.
We then show that this framework helps design better recommendation algorithms and make sense of their empirical performance through an online experiment.
Finally, we derive robust recommendation algorithms that achieve maximin performance from limited training data under varying assumptions about the compliance behavior of the decision-maker.
While our theorems and propositions explicitly hold for this stylized setup, the framework we develop is broadly applicable for designing algorithms that assist human decision-makers.
Specifically, we connect algorithm design choices to assumptions adopted from the instrumental-variables literature, exposing the central role of the reaction of the human decision-maker to algorithmic recommendations.

In \autoref{ms-sec:setup}, we introduce a \emph{causal design problem with endogenous response} that formalizes the objective of AI recommendation systems that assist a human decision-maker in making treatment choices.
In this model, an algorithm sends a recommended action to a decision-maker.
The decision-maker then chooses a binary action, which we also call the treatment, before some outcome of interest is realized.
For example, the decision-maker could be a doctor that receives an algorithmic recommendation of whether to test a patient or not.
Or she could be a judge who reacts to the output of a risk assessment tool.
Or the decision-maker could be a hiring manager who decides whether or not to employ an applicant with the help of an algorithmic hiring recommendation.
In all of these cases, we assume that the designer's goal is to maximize some beneficial outcome such as the patient's health, public safety, or the firm's productivity.
We allow the decision-maker to have additional, private information that the algorithm does not see, and
we also allow for complementarity in information and expertise between the algorithm and the decision-maker, thus moving beyond models in which perfect adherence to the algorithmic recommendation is optimal.
Our approach builds upon a classical treatment choice or policy learning problem where the algorithm directly chooses the action rather than leaving it to the human decision-maker.

Relative to a standard policy learning problem where an algorithm directly implements treatment choices, solving our algorithmic design problem requires accounting for endogenous decisions by the human decision-makers to the recommendations.
\autoref{ms-sec:po} therefore connects our problem setup to the instrumental-variables framework from causal inference, where we use the potential-outcome notation to express how decisions vary with different recommendations.
Using our ``potential decisions'', we can state the optimal recommendation algorithm as maximizing an objective that separates into two parts, (1) the expected outcome under a direct implementation of the algorithm's recommendations and (2) the gain (or loss) due to how decision-makers deviate from the recommendations.
We then introduce a monotonicity assumption that allows us to describe the effect of recommendations in terms of compliance, similar to the seminal analysis of \cite{imbens_identification_1994} for instrumental variables. This monotonicity assumption simplifies the analysis of endogenous responses since it implies that recommendations can only move the decision-maker toward the recommended action.
It also allows us to characterize human choices by a combination of their active decision they would take absent a recommendation as well as their compliance behavior.
As a result of this simplification, we can state our optimization problem in terms of a functional of active decisions and compliance alone, further linking the design of recommendations to the choice of optimal instruments.

In \autoref{ms-sec:experiment}, we showcase our framework by applying it to an online experiment that involves a hypothetical hiring task. This experiment asks subjects to make 25 hiring decisions and randomly varies the recommendation algorithm across subjects.
This setup allows us to evaluate how the structure of recommendation algorithms impacts ultimate task performance.
Specifically, the separate recommendation algorithms are based on different assumptions about compliance behaviors.
In line with our framework, we find that subjects perform better when the recommendations are designed to provide complementary information, rather than optimized for hiring decisions directly.
The best-performing algorithm assumes that human decision-makers comply with the algorithm selectively, only impacting the cases where the human is uncertain.
The decisions assisted by this algorithm outperform unassisted human decisions as well as the best performance by an algorithm on its own.
We thus demonstrate that our framework can help design systems that improve decisions through human--AI complementarity.

Finally, in \autoref{ms-sec:estimating}, we leverage our potential-decision framework to find robust recommendation algorithms from limited data.
Here, we map different assumptions about available data and potential decisions to maximin optimal recommendations.
We first focus on the realistic case where the designer may only have data about unassisted human decisions available and thus has to rely on assumptions about compliance and active decisions.
Assuming that the decision-maker is more likely to comply in cases where they do not make good decisions themselves, we show that learning successful recommendations is about \emph{learning from mistakes} rather than from the relationship of outcomes to characteristics alone.
We then consider an alternative approach that is motivated by the connection to instrumental variables:
an intent-to-treat (ITT) strategy based on randomized recommendations.
Across the approaches, we end by discussing the implementation of the resulting robust optimization program using machine learning.

Our work responds to an empirical literature that analyzes inefficiencies in the interactions between human decision-makers and prediction algorithms. This literature has identified cases where human decision-makers imperfectly use prediction algorithms in high-stakes decisions in criminal justice \citep{imai_experimental_2021,stevenson_algorithmic_2022,angelova_algorithmic_2023}, healthcare \citep{agarwal_combining_2023,maron_artificial_2020}, and child protective services \citep{fogliato_case_2022,mills_impact_2022}, so much so that the assisted human decisions are often no better than unassisted human decisions. A more extensive review of this literature is detailed by \cite{lai_towards_2021}. Of the works in this literature, our empirical approach is most related to that of \cite{green_principles_2019}, which posits ways human decision-makers should respond to algorithms and tests these hypotheses in the lab.
Relative to this literature, we focus on designing complementary recommendations, rather than focusing on describing ways in which humans misuse algorithms.

We join a collection of works spanning the areas of operations, economics, and computer science that design algorithms to assist humans in making better decisions. \cite{bansal_beyond_2019} and \cite{donahue_human-algorithm_2022} propose improvements to collaborative decision processes based on the private information of each agent. \cite{balakrishnan_improving_2022} and \cite{orfanoudaki_algorithm_2022} find that humans often incorrectly assume linear relationships and then develop strategies to help human--algorithm decision systems overcome this issue. \cite{athey_allocation_2020} and \cite{mclaughlin_algorithmic_2022} model how effort costs and reference points, respectively, affect a human's adoption of algorithmic assistance. \cite{snyder_algorithm_2023} shows how workers' compliance with an algorithm's recommendation in a service system varies with the system's utilization. \cite{baek_policy_2023} and \cite{bastani_improving_2022} apply ideas from reinforcement learning to incorporate past human actions into the recommendations of assistive algorithms. \cite{caro_believing_2023} and \cite{sun_predicting_2022} use data on workers' deviations from algorithm advice to increase compliance. \cite{noti_learning_2023} focuses on when to provide assistance, and shows that strategically restricting advice can preserve the complementary strengths of the algorithm and human. Our approach differs from these prior works as it provides a causal framework to capture how a human's response varies with the underlying algorithm. Similar to the work of \cite{ibrahim_eliciting_2021} (where human predictors assist algorithmic decision-makers), our results highlight the importance of private information in designing impactful collaboration.

Our approach to algorithmic recommendations relates to approaches in causal inference and algorithmic decision-making.
First, we build upon policy learning and statistical treatment choice \citep{manski_statistical_2004,kitagawa_who_2018,athey_policy_2021}.
Second, we relate to instrumental-variables approaches in cases with imperfect compliance \citep{imbens_identification_1994,angrist_identification_1996,vytlacil_independence_2002}, including the design of treatment rules with an instrument \citep{cui_semiparametric_2021,qiu_optimal_2021,pu_estimating_2021}.
Third, we connect to marginal treatment effects (MTE) and the related approaches to policy-relevant treatment effects \citep{heckman_structural_2005,mogstad_using_2018,mogstad_identification_2018}.
Fourth, we connect to the literature on selective labels and learning from human decisions \citep{lakkaraju_selective_2017,kleinberg_human_2017,rambachan_identifying_2024}.
Fifth, we join work by \citet{imai_experimental_2021} and \citet{ben-michael_does_2025} in leveraging potential outcomes for a causal analysis of algorithm-assisted human decisions.

\section{\texorpdfstring{A Causal Model of Algorithmically Assisted Decision-Making}{A Causal Model of Algorithmically Assisted Decision-Making}}
\label{ms-sec:setup}

We introduce a model of joint human--algorithm decision-making%
\begin{nonblind}, which builds upon our earlier work in \cite{mclaughlin_algorithmic_2022}\end{nonblind}%
. In this model, the designer chooses an algorithm that provides recommendations to a human decision-maker. The decision-maker observes each recommendation and subsequently makes a binary treatment choice.
We model outcomes of this choice using the standard potential-outcomes framework from causal inference.

\subsection{\texorpdfstring{Treatment Choice and the Automation Benchmark}{Treatment Choice and the Automation Benchmark}}

We consider the problem of taking a binary decision $D$, after which an outcome $Y$ is realized. We capture the impact of this decision through a utility function $\mathcal{U}$, which describes the (scalar) evaluation $\mathcal{U}(Y)$ of obtaining the outcome $Y$ achieved from implementing the decision $D$.
Since we are ultimately interested in maximizing expected utility \(\E[\mathcal{U}(Y)]\), we can assume without loss of generality that $Y$ is scalar and that $\mathcal{U}$ is the identity, so that the outcome of interest is already measured in utility terms and the optimization target is \(\E[Y]\).\footnote{For a general utility function \(\mathcal{U}\), all results below apply to the potential outcomes \(\mathcal{U}(Y(1))\) and \(\mathcal{U}(Y(0))\) in place of \(Y(1)\) and \(Y(0)\). We therefore drop \(\mathcal{U}\) from the notation.}
For each problem instance, we assume that there are baseline characteristics $X$ and $U$ that have predictive power for the values $Y$ takes for different choices of the treatment $D$.

We employ standard potential-outcomes notation from causal inference in order to express the effect of taking the decision $D \in \{1,0\}$ on the outcome $Y$. Specifically, we assume that there are potential outcomes $Y(1),Y(0)$ representing the consequence of making the choices $D{=}1$ and $D{=}0$, respectively. For the actual decision $D$, the outcome is then realized as $Y = Y(D)$%
\footnote{Throughout, the Stable Unit Treatment Value Assumption (SUTVA) that the outcome for one instance only depends on its own treatment is implicit to our framework.}
.
The potential outcomes $Y(1), Y(0)$ along with the covariates $X$ and $U$ follow a joint distribution $\P$.

The causal effect of treating a specific instance is $Y(1) - Y(0)$, and an optimal choice would be to choose $D = 1$ if and only if this difference is non-negative. However, we assume that the potential outcomes may be ex-ante unknown.
If we knew both the features $X$ available to the algorithm and the human information $U$, then the optimal (oracle) decision that maximizes expected utility would therefore be given by $D^* = \Ind[\tau^*(X,U) \geq 0]$, where $\tau^*(X,U) = \E[Y(1) - Y(0)|X,U]$ is the treatment effect conditional on both $X$ and $U$.
Since $U$ is not generally available to the algorithm, a standard policy learning approach would be to instead estimate the conditional average treatment effect (CATE) \[ \tau(X) = \E[Y(1) - Y(0) \mid X] \] and implement the rule \[ \fdecision \in \argmax_{\pi: X \mapsto \{0,1\}} \E[Y(\pi(X))], \qquad \fdecision(x) = \Ind[\tau(x) \geq 0], \] corresponding to the optimal direct choice by an algorithm that automates the decision. We refer to \(\fdecision\) as the \emph{automation benchmark}; it is the rule that standard policy learning targets.

\begin{exm}[Binary classification and hiring]
\label{exm:classificationhiring}

    As an example, consider a case where choosing treatment ($D{=}1$) comes with a cost of $c \in (0,1)$ and leads to a binary benefit $Y^* \in \{1,0\}$, relative to a net-zero outside option for choosing control ($D {=} 0$).
    For example, $D$ could represent a decision of whether to hire ($D{=}1$) or not hire ($D{=}0$) an applicant at cost $c$ with latent ability $Y^*$ that is good ($Y^*{=}1$) or bad ($Y^*{=}0$).
    Here, we assume that the latent outcome $Y^*$ is unaffected by treatment and unknown at decision time.
    The potential outcomes $Y(1),Y(0)$ associated with an instance with latent outcome $Y^*$ are then the net benefits
    \begin{align*}
        Y(1) &= Y^* - c,
        &
        Y(0) &= 0.
    \end{align*}
    The optimal choice of treatment in this case compares expected benefit to cost, $D^* = \Ind[\mu^*(X,U) \geq c]$ for $\mu^*(X,U) = \E[Y^*|X,U]$.
    It depends only on the oracle prediction $\mu^*(X,U)$ of the (latent) outcome $Y^*$, and thus takes the form of a \emph{prediction policy problem} \citep{kleinberg2015prediction}.
    This treatment choice can also be seen as a binary classification problem.
    Specifically, maximizing expected benefit
    is equivalent to minimizing the expected classification loss $\E[\ell(Y^*,D)]$ with the loss function
    \begin{align*}
        \ell(Y^*,D) = c_I \: \Ind[Y^*=1,D=0] + c_{II} \: \Ind[Y^*=0,D=1]
    \end{align*}
    for costs $c_I = 1- c, c_{II} = c$.
    Here, the type-I error cost $c_I > 0$ captures the relative cost of choosing $D=0$ for $Y^* = 1$ (e.g., the company not hiring a good worker). The type-II error cost $c_{II} > 0$ captures the relative cost of choosing $D=1$ for $Y^* = 0$ (hiring a bad worker).
    The choice that minimizes expected loss is then $D^* = \Ind\left[\mu^*(X,U) \geq \frac{c_{II}}{c_{I} + c_{II}}\right] = \Ind\left[\mu^*(X,U) \geq c \right]$ for $\mu^*(X,U) = \P(Y^*{=}1|X,U) = \E[Y^*|X,U]$, as before.

\end{exm}

\subsection{\texorpdfstring{Algorithmic Assistance with Endogenous Response}{Algorithmic Assistance with Endogenous Response}}

Above, we have introduced the problem of assigning binary treatment $D$ based on information $X,U$.
We now assume that the decision $D$ is taken by a human decision-maker with limited information and potentially misaligned preferences. Under assistance, the designer chooses a recommendation rule \[ f: X \mapsto R \in \{0, \Neutral, 1\}, \] where \(R = 0\) and \(R = 1\) recommend the two treatments and \(R = \Neutral\) stands for no recommendation, and the human decision-maker observes \(R\) and takes the decision \(D\). The objective \(\max \E[Y(D)]\) is unchanged. What changes is that the designer controls the decision only through the recommendation.
Specifically, we assume the decision $D$ is taken by a decision-maker who (1) does not observe the characteristics $X$ directly, (2) may observe private information determined by $U$, (3) may not share the designer's objective.
We allow for arbitrary overlap between the information in $X$ and the information observed by the decision-maker, so the model allows for none, some, or even all of the characteristics $X$ to be available to the decision-maker.
In causal terms, \(U\) is an unobserved confounder; it enters both the decision and the potential outcomes.

To improve the choices by the decision-maker, the designer implements an algorithm $f$ that provides recommendations $R = f(X)$ to the decision-maker.\footnote{We take \(f\) to be deterministic for simplicity. A natural extension considers randomized rules \(R = f(X, \xi)\), with noise \(\xi\) independent of the instance.}
Thus, similar to the information design literature \citep{kamenica_bayesian_2011,bergemann_information_2019}, the designer (as the sender) has partial control over the information presented to the decision-maker (as the receiver) before the latter makes the final decision.  We think of the algorithm as provided by an AI system that may learn from training data, but we abstract from the training process and focus on the resulting recommendation rule $f$ itself.

\begin{figure}[t]
    \centering

\begin{tikzpicture}[
  circle node/.style={draw, circle, minimum size=1.1cm, font=\small, align=center},
  >={Latex[length=2.5mm, width=2mm]},
  shaded/.style={circle node, fill=gray!30}
]
  \node[circle node] (X) at (3, -1) {$X$};
  \node[shaded] (U) at (5, -1) {$U$};
  \node[circle node] (Y) at (6, 1) {$Y$};
  \node[circle node] (R) at (2, 1) {$R$};
  \node[circle node] (D) at (4, 1) {$D$};
  \draw[->] (X) -- node[below left] {$f$} (R);
  \draw[->] (R) -- (D);
  \draw[->] (U) -- (D);
  \draw[->] (U) -- (Y);
  \draw[->] (D) -- (Y);
  \draw[->] (U) -- (X);
\end{tikzpicture}
    \caption{This diagram illustrates how the elements of our model are related to each other. The algorithm's information $X$ determines the recommendation $R$ through the rule $f$. The decision $D$ responds to the recommendation and to the private information $U$ of the decision-maker. The outcome $Y$ depends on the decision and on $U$. The shading of $U$ illustrates that the decision-maker's private information is never observable to the algorithm directly.}

    \label{ms-fig:diagramcausal3}
\end{figure}

\begin{exm}[continues=exm:classificationhiring]

In the hiring example, we assume the designer is a company that implements an algorithmic recommendation system $f$ that uses the characteristics $X$ on workers to generate hiring recommendations $R = f(X)$, which influence the hiring decisions of the decision-maker, who is a manager at the company. The recommendations reduce the unstructured applicant data $X$ to a simpler form that may still be informative about the worker's ability $Y^*$.
For simplicity, we limit the algorithm to either recommend a decision to the manager (not hire, $R =0$, or hire, $R =1$) or not make a recommendation at all ($R = \Neutral$).
In practice, recommendations may present more granular predictive information, such as a score that represents a scaled probability that a worker is good ($Y^*=1$) and allows the managers to understand the algorithm's relative ranking of different workers. It may also include an explanation of the score, highlight pieces of $X$ the algorithm deems particularly relevant, or include certainty information about the confidence of its prediction \citep{raghavan_mitigating_2020}.

The manager responds to the recommendation by making a hiring decision $D$ which may depend on the recommendation $R$ along with their private information $U$ and unknown private preferences. This information $U$ consists of all knowledge the manager has of the applicant, including potentially some of the information contained in $X$ (such as the candidate's resume) as well as new information the manager learns through their own conversations with the applicant (such as prior relationships with employees at the company or interest in areas the company is looking to expand). Thus, we allow arbitrary correlation between $X$ and $U$, but do not allow the manager to observe $X$ directly. This effectively captures the idea that $X$ and $U$ contain shared information, but that neither necessarily dominates the other. Additionally, the manager may exhibit preferences that do not directly align with maximizing the outcome $Y$, preferring to hire individuals with whom they get along well (or have prior relationships) even when those applicants are less likely to be good than others. This flexibility allows us to capture the complex decisions managers exhibit empirically. We summarize the relationship between the elements of our model in \autoref{ms-fig:diagramcausal3}.

\end{exm}

Throughout, we include the option that the algorithm does not make a recommendation, $R=\Neutral$, forcing the decision-maker to make an active decision.
We believe that this design feature has formal advantages for modeling recommendations and practical relevance for achieving human--algorithm complementarity.
First, we can consider unassisted decisions, those taken without any recommendation algorithm, as equivalent to those taken under a recommendation algorithm which always sends $R = \Neutral$, making unassisted decisions a special case of our framework.
Second,
\begin{nonblind}
building upon our earlier work in \citet{mclaughlin_algorithmic_2022},
\end{nonblind}
forcing active decisions by
avoiding recommendations in specific cases
may reduce distortions in cases where human decision-makers know better. These beliefs are further supported by the work of \cite{noti_learning_2023}, which finds that partially withholding algorithmic assistance (i) increases humans' reliance on advice when given and (ii) can be leveraged strategically to improve humans' prediction accuracy.

\section{Making Sense of Recommendations with Potential Decisions}
\label{ms-sec:po}

Having set up a treatment choice problem with endogenous responses for thinking about the design of recommendation algorithms, we now expand the potential-outcomes framework from causal inference to model how human decisions vary with recommendations. Viewing the decision-maker's decision in this way suggests that the descriptive and prescriptive analysis of recommendations shares common features with the investigation of instrumental variables in inference problems. We strengthen this connection by applying a monotonicity assumption similar to that of \cite{imbens_identification_1994}, which allows us to represent the remaining potential decisions in terms of the active decision the decision-maker makes when no recommendation is given and whether the decision-maker complies with the recommendation. This leads to a decomposition of the recommendation algorithm's objective based on the decision-maker's compliance choice. The objective captures the performance of the recommended action if the decision-maker complies and the performance of the decision-maker's active decision otherwise.

\subsection{\texorpdfstring{Recommendations Induce Potential Decisions}{Recommendations Induce Potential Decisions}}
\label{ms-sec:potdec}

We now interpret the decision-maker's decision in terms of the potential-outcomes framework of \cite{rubin_estimating_1974}.
Specifically, we write $D_{f}(r)$ for the ``potential decisions'' that the decision-maker would take when receiving the recommendation $r$ from the recommendation algorithm $f$ when the decision-maker's information is $U$. Thus, the decision-maker's decisions can be seen as the random variable $D = D_{f}(R)$, where $R = f(X)$.
The subscript \(f\) records that the response to a given recommendation may depend on the rule that produces it; we return to this dependence in \autoref{ms-sec:dependence}.

Since the decision-maker's information $U$, the recommendation $R$, and the algorithm $f$ are sufficient information to determine $D$, and the designer controls the recommendation algorithm $f$ (and through it $R$), the decision is the potential decision $D_{f}(r)$ for each of the possible values $r \in \{0,\Neutral,1\}$ of the recommendation. Since decisions are themselves binary and the recommendations take three values, we can enumerate these potential decisions into eight triplets which we categorize into four response types based on how the realized recommendation impacts the assisted decision-maker's decisions. These response types are similar in nature to the types developed by \citet{angrist_identification_1996} for the analysis of instrumental variables.

\autoref{ms-tab:potentialdecisions} enumerates the potential choices associated with a decision-maker's response to recommendations and provides the aforementioned categorization. If the decision-maker maintains the same decision regardless of the recommendation, they ignore the recommendation (similar to always-takers and never-takers in causal inference). If they adopt the recommendation regardless of whether the algorithm recommends not hiring (0) or hiring (1) the applicant, then they comply with the recommendation. If, instead, the decision-maker always takes the decision that the algorithm does \emph{not} recommend, then they defy the recommendation. If none of these cases hold, then it must be the case that the decision-maker will change their decision in response to the algorithm sending a recommendation, taking one decision when the algorithm sends no recommendation and taking the opposite decision if either recommendation is sent. While we maintain all these response types for now, we will later simplify the analysis by ruling out the defy and change types, analogous to the monotonicity assumption of \citet{angrist_identification_1996}.

\begin{table}
\begin{center}
\begin{tabular}{cccc}
\toprule
Response type & $D_{f}(0)$ & $D_{f}(\Neutral)$ & $D_{f}({1})$ \\
\midrule
\multirow{2}{*}{\centering Ignore} & 0 & 0 & 0 \\
& 1 & 1 & 1\\
\midrule[0.5pt]
\multirow{2}{*}{\centering Comply} & 0 & 0 & 1 \\
 & 0 & 1 & 1 \\
 \midrule
\multirow{2}{*}{\centering Defy} & 1 & 0 & 0 \\
& 1 & 1 & 0 \\
\midrule[0.5pt]
\multirow{2}{*}{\centering Change} & 1 &0 & 1\\
& 0 & 1 & 0 \\
\bottomrule
\end{tabular}
\caption{Enumeration and categorization of the potential decisions of the decision-maker in response to different realizations of the recommendation. The defy and change response types will later be ruled out by \autoref{ms-asm:Mon}.}
\label{ms-tab:potentialdecisions}
\end{center}
\end{table}

\subsection{\texorpdfstring{Recommendations as Instruments}{Recommendations as Instruments}}
\label{ms-sec:iv}

In the language of causal inference, our framework is an instrumental-variables problem: the recommendation is an instrument that affects the outcome only through its impact on the treatment, here the decision.
In our hiring example, the algorithm's recommendations (corresponding to the instrument) impact the company's utility (playing the role of the outcome) only through managers' decisions (treatment).
Rather than directly optimizing for the recommendation, the designer must understand how the decision-maker's decisions $D$ vary in response to receiving different recommendations from different algorithms.

The condition that makes the recommendation an instrument in the sense of \citet{angrist_identification_1996} is already built into the notation of \autoref{ms-sec:setup}, where the potential outcomes are indexed by the decision alone; we state it as an assumption to make visible what the notation rules out.

\begin{asm}[Exclusion]
\label{ms-asm:exclusion}
For every recommendation rule $f \in \F$, every decision $d \in \{0,1\}$, and every recommendation $r \in \{0, \Neutral, 1\}$, the potential outcome of the decision does not depend on the recommendation: writing $Y_f(d, r)$ for the outcome that would be realized under rule $f$, recommendation $r$, and decision $d$, we have $Y_f(d, r) = Y(d)$.
\end{asm}

\noindent Exclusion fails when the recommendation changes how a decision is carried out, or when the human's action has dimensions that the binary $D$ does not record, such as effort or timing. Since $R = f(X)$, the recommendation is independent of $(Y(1), Y(0), D_f(\cdot))$ conditional on $X$ but not unconditionally, so all instrumental-variables statements in this paper are conditional on $X = x$, with $U$ in the role of the unobserved confounder of $D$ and $Y$. What is \emph{not} assumed is that the potential decisions $D_f(\cdot)$ are the same across rules $f$; \autoref{ms-sec:dependence} returns to this point. \autoref{ms-fig:diagram-iv} shows the structure.

\begin{figure}
    \centering

\begin{tikzpicture}[
  circle node/.style={draw, circle, minimum size=1.1cm, font=\small, align=center},
  >={Latex[length=2.5mm, width=2mm]}
]
  \node[circle node, draw=lightgray, text=lightgray] (X) at (3, -1) {$X$};
  \node[circle node, draw=lightgray, fill=gray!10, text=lightgray] (U) at (5, -1) {$U$};
  \node[circle node] (Y) at (6, 1) {$Y(D)$};
  \node[circle node] (R) at (2, 1) {$R$};
  \node[circle node] (D) at (4, 1) {$D(R)$};
  \draw[->, lightgray] (X) -- node[below left, lightgray] {$f$} (R);
  \draw[->] (R) -- (D);
  \draw[->, lightgray] (U) -- (D);
  \draw[->, lightgray] (U) -- (Y);
  \draw[->] (D) -- (Y);
  \draw[->, lightgray] (U) -- (X);
\end{tikzpicture}
    \caption{The recommendation as an instrument. The recommendation $R$ affects the outcome only through the decision (\autoref{ms-asm:exclusion}) and is assigned as a function of the algorithm's information $X$; the private information $U$ confounds the decision and the outcome.}
    \label{ms-fig:diagram-iv}
\end{figure}

\subsection{\texorpdfstring{Monotonic Responses, Compliance, and Active Decisions}{Monotonic Responses, Compliance, and Active Decisions}}
\label{ms-sec:mon}

Above, we have proposed a potential-outcomes framework for human decisions in response to recommendations.
We now introduce assumptions that limit the potential decision to simplify their analysis.
We then use these simplifications to shed further light on the structure of the optimal recommendation algorithm in terms of compliance and active decisions.

In the context of the four response types introduced in \autoref{ms-sec:potdec} and listed in \autoref{ms-tab:potentialdecisions},
behavior consistent with the ignore and comply responses is easier to make sense of than behavior consistent with the defy and change responses. If the manager believes the algorithm made a bad recommendation given their private information $U$ about the applicant, they are likely to ignore the recommendation. Conversely, if they believe the algorithm makes a good recommendation given their private information $U$ about the applicant, they will comply. Using recommendations as a threshold requirement for considering an applicant, as done by hiring managers interviewed by \citet{li_algorithmic_2021}, also generates potential outcomes consistent with only ignore and comply responses. On the other hand, defy and change behaviors are counter-intuitive and typically need bad recommendation algorithm designs, perverse correlation between $X$ and $U$, or adversarial managerial incentives to be justified rationally.
Consequently, we now make an assumption to eliminate the potential outcomes associated with defy and change responses, similar to the monotonicity assumption of \citet{angrist_identification_1996} in the context of instrumental variables and its extensions to multivariate discrete treatments \citep[including][]{lee2018identifying}.

\begin{restatable}[Monotonic response to $R$]{asm}{mon}
    \label{ms-asm:Mon}
    Let $\preceq$ be the ordering of decisions with $0 \preceq 1$. We assume, for all $f \in \F$, that $D_{f}(\cdot)$ is such that $D_{f}(0) \preceq D_{f}(\Neutral) \preceq D_{f}(1)$.
\end{restatable}

\noindent At a high level, this assumption restricts the direction of the decision-maker's response to follow the algorithm's recommendation. In the hiring example, hire recommendations ($R =1$) can turn not-hire decisions into hire decisions ($D = 0 \to D = 1$) but cannot turn hire decisions into not-hire decisions ($D = 1 \not\to D = 0$). Not-hire recommendations ($R = 0$) do the exact opposite.

The monotonicity restriction exactly rules out the bottom four rows of \autoref{ms-tab:potentialdecisions}.
It can be considered a restriction on plausible recommendation rules $\F$.
Specifically, it rules out that recommendation rules are used that would make sense to defy (such as a rule that suggests hiring bad candidates) or that would make sense to observe the change behavior for (such as a rule that only gives a recommendation for bad candidates and never gives one for good candidates).

When the decision-maker's behavior is limited by \autoref{ms-asm:Mon}, we can fully describe their behavior for some specific recommendation rule $f$ in terms of their active decisions $\Dactive = D_{f}(\Neutral)$ which they exhibit in response to the algorithm $f$ being implemented, along with their compliance behavior $C$:

\begin{restatable}[Simplified potential outcomes]{prop}{spo}
    \label{ms-prop:simpPO}
    Under \autoref{ms-asm:Mon}, the potential outcome triplet $(D_{f}(0),D_{f}(\Neutral),D_{f}(1))$ can be fully expressed by the active decision $\Dactive_f = D_{f}(\Neutral) \in \{0,1\}$, which expresses the decision-maker's decision when not given a recommendation, and the compliance type $C_f \in\{0, 1\}$, which specifies whether the decision-maker adopts $R=f(X)$ when the recommendation is presented.
\end{restatable}

\noindent \autoref{ms-tab:potential_outcomes} visualizes \autoref{ms-prop:simpPO} by restating the remaining potential-decision types in terms of $(\Dactive,C)$ once \autoref{ms-asm:Mon} is applied. The only information determining the decision $D$ outside of the recommendation $r$ is the active decision $\Dactive$ and whether the response type $C$ of the decision-maker is $0$ (ignore) or $1$ (comply). The proposition follows directly from an inspection of \autoref{ms-tab:potentialdecisions}. For completeness, a formal proof is given in \autoref{ms-sec:ProofApp}.
Written out, the representation is \[ D = \begin{cases} R & \text{if } R \neq \Neutral \text{ and } C_f = 1, \\ \Dactive_f & \text{otherwise,} \end{cases} \] which is the identity that the decomposition of the objective in \autoref{ms-subsec:Objectives} uses.

\begin{table}
\begin{center}
\begin{tabular}{lccc}
\toprule
$C_f$ & $D_{f}(0)$ & $D_{f}(\Neutral)$ & $D_{f}({1})$ \\
\midrule
{\centering $0$} & $\Dactive_f$ & $\Dactive_f$ & $\Dactive_f$ \\
{\centering $1$} & 0 & $\Dactive_f$ & 1 \\
\bottomrule
\end{tabular}
\caption{Reduction of potential outcomes under \autoref{ms-asm:Mon}. This organization acts as a visual representation of \autoref{ms-prop:simpPO}, showing how the values that the potential outcomes take reduce to the active decision $\Dactive \in \{0{},1{}\}$ absent a recommendation and the compliance $C \in \{0, 1\}$.}
\label{ms-tab:potential_outcomes}
\end{center}
\end{table}

\subsection{\texorpdfstring{Dependence on the Rule}{Dependence on the Rule}}
\label{ms-sec:dependence}

The potential decisions $D_{f}(r)$ depend not only on possible recommendation values $r$ of $R$, they also depend on the whole mapping $f$ from characteristics $X$ to recommendations $R = f(X)$.
This is because how a \emph{specific} recommendation is interpreted depends on how recommendations are given overall.
For example, if an algorithm recommends hiring for \emph{everybody}, then it may change individual decisions differently from it recommending to hire only a few applicants.
This is related to the concept of \emph{interference} in causal inference, where potential outcomes may be functions not just of their own treatment assignment, but also of those of others.
Potential decisions may also depend on context: on how the rule is explained, on the interface, or on the decision-maker's experience with the algorithm. The rule \(f\) can then be read as the design as a whole.
This dependence highlights the challenge in estimating good recommendation algorithms under minimal assumptions or using limited data (which we return to in \autoref{ms-sec:estimating}).
Effective ways of forming recommendation algorithms will, therefore, rely on assumptions that limit the complexity of potential decisions, including their dependence on the overall recommendation algorithm.%
\footnote{
Without additional assumptions, this interference challenge implies, in our case, that the potential decision function $D_{f}(\cdot)$ for a \emph{specific} instance $(X,U)$ is only ever evaluated at its implied recommendation $R=f(X)$.
However, the decision-maker does not observe $X$, so $X$ affects $D$ only through $R$: for a given rule $f$, the potential decision function $D_{f}(\cdot)$ is a function of $U$ alone. This exclusion restriction at the decision stage is what makes $D_f(r)$ well defined for recommendations $r \neq f(X)$, and the notation then also captures variation in recommendations given for instances that are otherwise equivalent from the perspective of the human decision-maker.
That is, for a given instance $U$, they capture how decisions vary across possible realizations of $f(X)$ stemming from variation in $X$.
This formalization, therefore, implies that two instances that present with the same $U,R$ values are not distinguishable to the decision-maker, leading to the same decision even if their $X$ varies. (Here, any noise in decisions would be captured through $U$.)
}

\subsection{\texorpdfstring{Objectives of Recommendation Algorithms}{Objectives of Recommendation Algorithms}}
\label{ms-subsec:Objectives}

Having introduced a framework to capture potential decisions in response to recommendations,
we now consider the optimal recommendation algorithm as the solution to an optimization problem that maximizes the expected outcome achieved by the decisions the algorithm induces.
We show that the recommendation's objective can be separated into a triage term that assumes that recommendations are implemented directly and a term that captures the decision-maker's endogenous response to the recommendation. We compare this optimal recommendation algorithm to solutions that instead directly target optimal decisions, and provide an intuition for why the automation benchmark will rarely be the optimal recommendation rule.

In general, the optimal recommendation algorithm is the one that maximizes the expected outcome of the decision taken by the human decision-maker after seeing the recommendation,
\begin{equation}
f^* = \underset{f:X\mapsto R}{\mathrm{argmax}} \ \E\left[Y\right]  \text{ where } R = f(X), D = D_f(R), Y = Y(D).
\label{ms-eq:OptSol}
\end{equation}
In our hiring example, the optimal recommendation algorithm for the company to implement is the mapping from applicant worker data to recommendations that leads to good hiring decisions by the manager.
As this example illustrates, this optimization problem is generally quite difficult to solve, given the uncertainty about how managers will respond to recommendations. As a result, designers often implement algorithms that instead recommend the optimal decision based on the information available to the algorithm, the automation benchmark of \autoref{ms-sec:setup},
\begin{equation}
\fdecision = \underset{f:X\mapsto R}{\mathrm{argmax}} \  \E\left[Y(R)\right] \text{ where } R = f(X), R \neq \Neutral,
\label{ms-eq:OptDec}
\end{equation}
implicitly assuming that the recommendation $R=f(X)$ gets implemented directly ($D=R$) and ignoring the algorithm's ability to not send a recommendation ($R = \Neutral$).
An extension suggested in previous work by \cite{raghu_algorithmic_2019} is to triage instances into those where algorithms outperform humans and those where humans outperform algorithms.
We write $\Dnull = D_{f^{\Neutral}}(\Neutral)$ where $f^\Neutral \equiv \Neutral$ for the decision taken by the decision-maker in the absence of a recommendation algorithm. If we assume that the decision-maker takes this decision when not given a recommendation ($f(X) = \Neutral$) and that otherwise the recommendation is implemented directly, then this objective and its solution correspond in our framework to
\begin{equation}
\label{ms-eqn:triage}
\ftriage = \underset{f:X\mapsto R}{\mathrm{argmax}} \ \E\left[Y(R)\:\Ind\left[R\neq \Neutral\right] + Y(\Dnull)\:\Ind\left[R= \Neutral\right]\right] \text{ where } R = f(X).
\end{equation}
In practice, recommendations may not be followed perfectly. In addition, even the decision for $R = \Neutral$ may itself be different from the decision $\Dnull$ without a recommendation algorithm, and we may specifically hope it to improve since additional information and effort may be available for those cases \citep{raghu_algorithmic_2019,athey_allocation_2020}.

We now compare these different objectives and solutions, and show that the optimal recommendation ($f^*$) differs from triage ($\ftriage$) in that it also incorporates the compliance behavior of the decision-maker and considers how active decisions may differ from unassisted decisions. Specifically, the objective that defines the optimal recommendation can be decomposed into the objective of the triage problem (which we call the triage effect, \textit{TE}) and the response effect (\textit{RE}) from decision-makers deviating from the triage decision,
\begin{equation}
\label{ms-eqn:decomposition}
\begin{aligned}
    f^* = \underset{f:X\mapsto R}{\mathrm{argmax}} \  \ &\overbrace{\E\left[Y(R) \: \Ind\left[R\neq \Neutral \right] + Y(\Dnull) \: \Ind\left[R = \Neutral\right]\right]}^{\textnormal{\textit{TE} -- Triage Effect}}\\
    &+\underbrace{\E\left[\left(Y(D)-Y(R)\right) \: \Ind\left[R\neq \Neutral \right] + \left(Y(D)-Y(\Dnull)\right) \: \Ind\left[R = \Neutral\right]\right]}_{\textnormal{\textit{RE} -- Response Effect}}\\
    &\text{where $R = f(X), D = D_f(R), Y = Y(D)$.}
\end{aligned}
\end{equation}
In addition to the triage effect from \eqref{ms-eqn:triage}, this decomposition has a second part representing deviations from recommendations and unassisted decisions. It can represent both improvements (in cases where the human decision-maker uses their private information to override a suboptimal recommendation or adjusts their active decision based on the knowledge they receive through the recommendation algorithm) as well as mistakes (when the decision-maker goes against a correct recommendation or takes an active decision that is worse than in the case without any recommendations).

In the language of instrumental variables, the left-hand side of \eqref{ms-eqn:decomposition}, $\E[Y(D_f(R))]$, is the intent-to-treat value of the rule: the recommendation is assigned, and the human decides.

The two components also highlight the challenge of learning good recommendations from data.
The triage effect (\textit{TE}) is straightforward to estimate, provided the designer has access to historical data through which they can learn the joint distribution of characteristics $X$, potential outcomes $(Y(1), Y(0))$, and unassisted decisions $\Dnull$.\footnote{This is the case in the experiment of \autoref{ms-sec:experiment}, where the latent ability of every applicant is known. With only the realized outcome $Y = Y(\Dnull)$ in the data, the triage effect is not identified without further assumptions; \autoref{ms-sec:estimating} takes up this case.}
On the other hand, the response effect depends on the potential decisions of the decision-maker, which complicates its evaluation for any $f$ under which decision data has not already been collected.
Understanding $D_{f}(r)|X,Y(1),Y(0)$ may require more than simple experimental evaluation based on local changes in recommendations.
These decision distributions may vary arbitrarily with changes to $f$.
Solving for the optimal recommendation then requires understanding this distribution across a set of reasonable recommendation algorithms $\F \subseteq \{f; f:X \mapsto R\}$. This makes designing an optimal recommendation algorithm an intractable task without additional assumptions (even when we just want to robustly outperform the unassisted decisions $\Dnull$).

As an immediate consequence of the simple representation of response types in \autoref{ms-prop:simpPO}, we can decompose the objective of the recommendation algorithm in terms of compliance:
\begin{equation}
\label{ms-eqn:compliancedecomposition}
\begin{aligned}
    f^* = \underset{f:X\mapsto R}{\mathrm{argmax}} \  &\E\left[Y(R) \: \Ind\left[C = 1 \text{ and } R\neq \Neutral \right]
    + Y(\Dactive) \: \Ind\left[C = 0 \text{ or } R = \Neutral\right] \right]
    \\
    &\text{where $R = f(X)$}
\end{aligned}
\end{equation}
Similarly, this explicit notation allows us to shed further light on the response effect from \eqref{ms-eqn:decomposition} in terms of compliance and deviations between active decisions and unassisted decisions:
\begin{equation*}
    \textit{RE}
    =
    \E\left[\left(Y(\Dactive)-Y(R)\right) \: \Ind\left[R\neq \Neutral,C=0 \right] + \left(Y(\Dactive)-Y(\Dnull)\right) \: \Ind\left[R = \Neutral\right]\right]
\end{equation*}
The first term in this decomposition expresses deviations from recommendations when they are ignored, while the second term captures the change in behavior for those instances where no recommendation is given that comes from the presence of the recommendation algorithm.

The same representation gives the objective in terms of treatment effects. With $\tau = Y(1) - Y(0)$, \autoref{ms-prop:simpPO} implies $Y(D) = Y(\Dactive_f) + (Y(R) - Y(\Dactive_f)) \, C_f \, \Ind[R \neq \Neutral]$, and since $Y(1) - Y(\Dactive_f) = (1 - \Dactive_f) \tau$ and $Y(0) - Y(\Dactive_f) = -\Dactive_f \tau$,
\begin{equation}
\label{ms-eqn:complier}
    \E[Y(D_f(R))] = \E[Y(\Dactive_f)] + \E\bigl[\tau \, (1 - \Dactive_f) \, C_f \, \Ind[R = 1]\bigr] - \E\bigl[\tau \, \Dactive_f \, C_f \, \Ind[R = 0]\bigr].
\end{equation}
The value of a rule is the value of the active decisions, plus the treatment effects of the instances that a hire recommendation moves into treatment, minus the treatment effects of the instances that a not-hire recommendation moves out of it. What matters for the recommendation at $x$ is therefore the average effect among the compliers whose decision it changes, $\E[\tau \mid X = x, \Dactive_f = 0, C_f = 1]$ for a hire recommendation and $\E[\tau \mid X = x, \Dactive_f = 1, C_f = 1]$ for a not-hire recommendation, and not the conditional average treatment effect $\tau(x)$ that defines the automation benchmark.

\subsection{\texorpdfstring{Optimal Recommendations under Specific Assumptions}{Optimal Recommendations under Specific Assumptions}}
\label{ms-sec:special}

Even with the simplifications introduced by \autoref{ms-asm:Mon}, it is not generally possible to learn the optimal recommendation algorithm from past data $(Y,X,\Dnull)$ on outcomes, characteristics, and unassisted human decisions.
This is because the joint distributions of active decisions and compliance with outcomes and characteristics, $(Y,X,\Dactive,C)$ cannot be identified from such limited data without further assumptions.
However, we can now formulate conditions on these unobserved quantities to derive optimal learnable recommendation algorithms in specific cases, thus providing a formal analog and extension of the discussion in \autoref{ms-subsec:Objectives}:
\begin{itemize}
    \item If compliance is perfect ($C \equiv 1$) and decisions in instances without recommendations are not affected ($\Dactive \equiv \Dnull$), then the optimal recommendation algorithm is the optimal triage solution $\ftriage$ (which can be learned from data on $(Y,X,\Dnull)$).
    \item If compliance is perfect ($C \equiv 1$), and recommendations are limited to be binary ($R \neq \Neutral$) or human decisions are dominated by recommended actions (such as when $\E[Y(\Dactive)|X] \leq \max_{r \in \{1,0\}}\E[Y(r)|X]$ for all $f \in \F$), then the optimal recommendation algorithm is the automation benchmark $\fdecision$ (which can be learned from data on $(Y,X)$). This applies, specifically, in the case with perfect compliance and no private information of the human decision-maker.
\end{itemize}
At the same time, more nuanced assumptions or data on compliance behavior may be necessary to achieve complementarity in practical cases.
This may include the cases below:
\begin{itemize}
    \item If the recommendation algorithm is independent of compliance and active decisions ($\Dactive \equiv \Dnull$), then the optimal recommendation can generally be seen as the optimal triage decision \emph{among compliers only},
    \begin{equation*}
    \underset{f:X\mapsto R}{\mathrm{argmax}} \  \E\left[Y(R)\:\Ind\left[R\neq \Neutral\right] + Y(\Dnull)\:\Ind\left[R= \Neutral\right]\middle|C=1 \right] \text{ where } R = f(X).
    \end{equation*}
    However, unlike the case of full compliance, we now have to learn about compliance behavior in addition to active decisions.
    \item Even if the decision-maker makes optimal decisions given their available information, we would require information beyond $(Y,X,\Dnull)$ since the optimal recommendation would also depend on the structure and information content of the private signal $U$.
\end{itemize}
In our experiment below, we combine information about the joint distribution of $(Y(1),Y(0),X,U)$ with assumptions on active decisions $\Dactive$ and compliance $C$ in order to solve for sensible recommendation algorithms.

\FloatBarrier
\section{Complementary Recommendations in an Experiment}
\label{ms-sec:experiment}

We now apply our framework to the design and analysis of recommendation algorithms in a controlled trial.
In a pre-registered online experiment, study subjects participate in an incentivized hiring game.%
\footnote{The pre-analysis plan for this experiment can be found at \href{https://www.socialscienceregistry.org/trials/11857}{www.socialscienceregistry.org/trials/11857}.}
This empirical application showcases our framework's ability to map realistic assumptions about decision-maker behavior to the design of practical recommendation algorithms that realize complementarity between humans and algorithms.

In the experiment, we collect hypothetical hiring decisions while varying the structure of the recommendation algorithm at the subject level.
We apply different assumptions on how subjects will behave within our framework to generate these recommendation algorithms. We find that subjects, on average, improve their decisions with algorithms whose recommendations are focused on providing suggestions that are complementary to the subjects' private information. Moreover, subjects given these focused recommendations perform better than subjects who see naive recommendations based on optimal algorithmic predictions and outperform decisions by the human or algorithm alone.
In the notation of the framework, the treatment $D$ is the hiring decision; the potential outcomes are $Y(1) = +1$ for a good applicant and $Y(1) = -1$ for a bad one, and $Y(0) = 0$, which is the hiring example of \autoref{ms-sec:setup} with $c = 1/2$ after rescaling; $X$ is the applicant's personality type, which only the algorithm sees, and $U$ is the applicant's role, which only the subject sees, together with any noise in the subject's decisions. Each treatment arm is a rule $f$, randomized across subjects, so that each arm identifies the value $\E[Y(D_f(R))]$ of its rule, and the share of correct decisions equals $\E[Y(D)]$ up to the constant $\P(Y^* = 0)$. Because the latent ability $Y^*$ is known for every applicant, both potential outcomes are known for every instance, which is the data condition that \autoref{ms-sec:estimating} calls observed outcomes.

\subsection{Experimental Setup}

We implement a hiring game as part of an online experiment to test subjects' performance when using recommendation algorithms in a setting where both the subject and algorithm have private information. Specifically, we label 25 hypothetical applicants as good or bad based on the role they are applying for and their personality type, and ask subjects to only hire good applicants.
The role is only known to the human subject and the personality type is only known to the algorithm.
We measure the change in the accuracy of subjects' hiring decisions in response to different recommendation algorithms (treatments) that vary at the subject level.

\begin{figure}
    \centering
    \includegraphics[width = \textwidth]{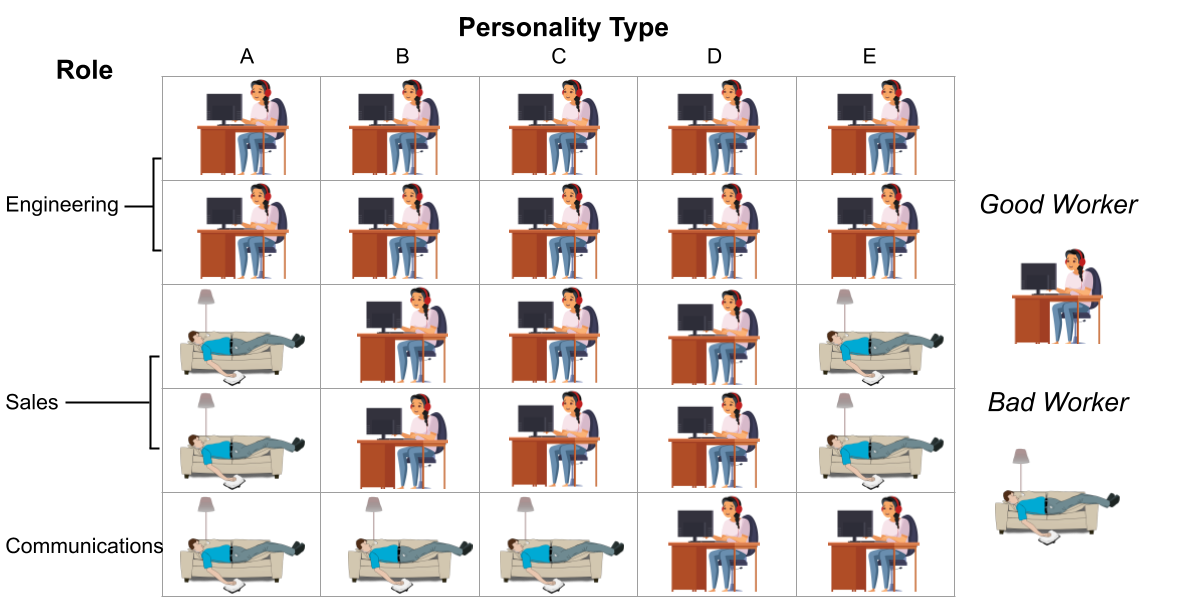}
    \caption{Summary of how the information available to the subject (the applicant's role) and the information of the algorithm (the applicant's personality type) relate to the applicant's ability (bad or good). This information is shown to all participants in the experiment and is easily accessible throughout their decision-making process.}
    \label{ms-fig:popinfo}
\end{figure}

For each subject, the experiment proceeds as follows. Subjects arrive at a Qualtrics survey from the recruiting site Prolific and learn they will make hiring decisions on 25 hypothetical applicants, of which there are good and bad types (referred to as the applicant's ability). They learn that each applicant's ability $Y^*$ is determined by the relationship of the applicant's personality type $X$ (denoted by letters A--E) to their intended role $U$ (Engineering, Sales, or Communications), as shown in \autoref{ms-fig:popinfo}. Then they answer three comprehension questions about how ability $Y^*$ is determined and, conditional on answering correctly, we randomize them into a treatment.

\begin{figure}
    \centering
    \includegraphics[width = 0.8\textwidth]{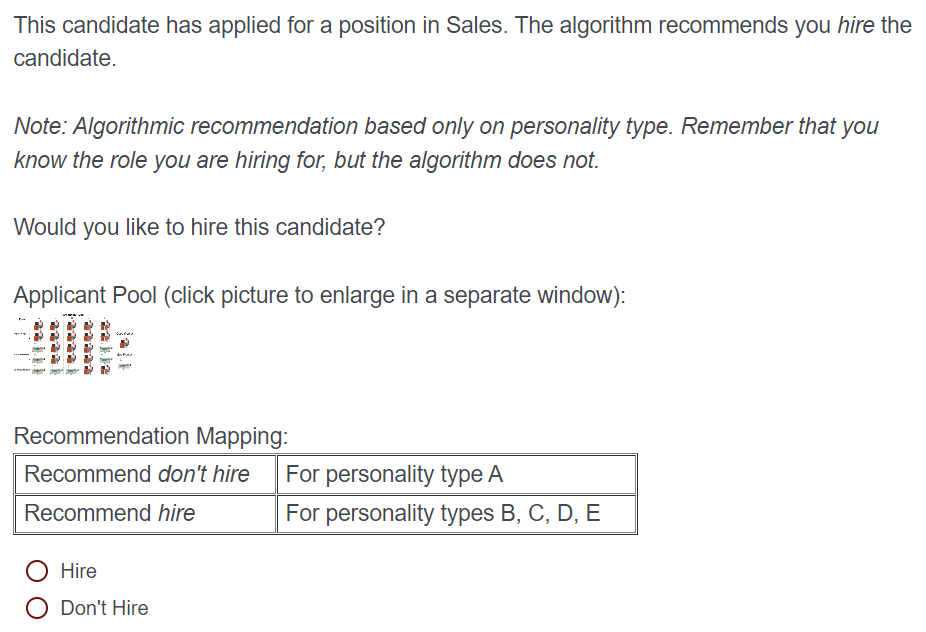}
    \caption{Sample hiring decision from the \textit{Predictive} treatment.}
    \label{ms-fig:exampleQ}
\end{figure}

Subjects receive recommendations from algorithms to assist them with making hiring decisions, except for those subjects in the control arm. Unlike the study subjects, these algorithms can access the applicants' personality types, but not their role.%
\footnote{Subjects then answer three more comprehension questions about the specific algorithm they will interact with, but continue in the experiment regardless of their answers. We leverage these additional questions for robustness checks below.}
Subjects then make 25 hiring decisions with access to the distribution of applicants as represented in \autoref{ms-fig:popinfo}, the recommendations of their assigned algorithm, and an explanation of that specific recommendation algorithm. An example of a hiring decision is given in \autoref{ms-fig:exampleQ}. When making these hiring decisions, subjects know the applicant's role, but are unaware of the applicant's personality type beyond any information conveyed by the recommendation.

We incentivize subjects to hire good applicants and to not hire bad applicants.
Specifically, as compensation for participating in our experiment, subjects receive \$2.04 for their time (estimated to be 12 minutes) and the chance to earn a \$2.00 bonus as a performance incentive. They earn this bonus if their hiring decision on a randomly selected applicant is correct (hire if good, not hire if bad). Note that this induces equal type I and type II losses ($c_I = c_{II}$) from the perspective of our subjects, and we analogously evaluate their performance using the percentage of correct decisions taken.
Subjects see all 25 applicants in a random order without replacement to make their hiring decisions but don't learn whether they made correct decisions.

The joint distribution of applicant role, personality type, and ability (pictured in \autoref{ms-fig:popinfo}) is designed to test subjects' willingness to comply with recommendations across applicant roles while allowing the recommendation to provide useful information in cases the applicant role is relatively uninformative. When subjects review an applicant for an Engineering role, they know the worker is good. Subjects are much more uncertain for Sales applicants (60\% of which are good) and Communications applicants (40\% of which are good). The algorithm's private information (personality type) varies significantly across the cases in which subjects are uncertain, thus containing useful information to help them make their decisions. Moreover, both the subject and algorithm's private information are of equal value in that a rational agent with access to either the role or the personality type would take the correct decision $76\%$ of the time.

\subsection{Treatments}
\label{ms-sec:treatdev}

The recommendation algorithm a subject receives represents the treatment in our experiment and is randomized at the subject level.
In addition to four treatment arms representing different recommendation algorithms, there is a control arm in which no recommendation is given.
These recommendation algorithms differ in the way they map personality types $X$ to recommended actions $R = f(X) \in \{0,\Neutral,1\}$, where the recommendations follow our earlier example and correspond to not hiring, no recommendation, and hiring the applicant, respectively.
We now describe the four different algorithms and link them to the implicit assumptions about compliance and active decisions that motivate their design. Additionally, \autoref{ms-tab:treatments} lists the recommendations that these algorithms provide across personality types.

\begin{table}
    \centering
    \begin{tabular}{lccccc}
        \toprule
        Algorithm & A & B & C & D & E \\
        \midrule
        \textit{Control} &   $\Neutral$ & $\Neutral$ & $\Neutral$ & $\Neutral$ & $\Neutral$ \\
        \textit{Predictive} &0 & 1 & 1 & 1 & 1 \\
        \textit{Complementary} &  0  & 1 & 1 & 1 & 0  \\
        \textit{Triage} &  0  & $\Neutral$ & $\Neutral$ & 1 & 1 \\
        \textit{Complementary Triage}  & 0  & $\Neutral$ & $\Neutral$ & 1 & 0  \\
        \bottomrule
    \end{tabular}
    \caption{Recommendations that each algorithm sends across applicants' personality types $X \in \{ \textnormal{A},\textnormal{B},\textnormal{C},\textnormal{D},\textnormal{E}\}$. Note that personality types B and C are equivalent for the distribution of outcomes as well as in terms of recommendations given.}
    \label{ms-tab:treatments}
\end{table}

\begin{description}
    \item[\textit{Control}.]
Subjects in this condition do not observe a recommendation algorithm. We use this treatment to collect information on the distribution of subjects' unassisted decisions.

\item[\textit{Predictive}.] The algorithm in this condition sends a hire recommendation if at least half of the applicants associated with the given personality type are good and sends a not-hire recommendation otherwise. This \textit{Predictive} algorithm is the optimal decision an algorithm can take by itself to minimize overall errors, so it corresponds to $\fdecision$ in the framework of \autoref{ms-sec:po}. It would correspond to the optimal \emph{recommendation} if compliance were perfect and active decisions only led to worse hiring decisions.

\item[\textit{Complementary}.] These subjects observe an algorithm that sends a hire recommendation if at least half of the Sales and Communications applicants associated with the given personality type are good and sends a not-hire recommendation otherwise. This algorithm ignores Engineering applicants on the assumption that subjects will perform effectively by themselves when asked whether to hire these applicants, and instead sends the optimal decision for the subpopulation of Sales and Communications applicants. It corresponds to the optimal recommendation if subjects do \emph{not} comply with the recommendation for Engineering candidates, but comply perfectly for the remaining applicants.

\item[\textit{Triage}.] These subjects observe a version of the \textit{Predictive} algorithm that implements a `safety check', causing the algorithm not to send recommendations for some personality types. Specifically, the \textit{Triage} algorithm only sends the \textit{Predictive} algorithm's recommendation of hire if at least half of the Communications applicants associated with the given personality type are good and it only sends the \textit{Predictive} algorithm's recommendation of not hire if at least half of the Engineering and Sales applicants associated with the given personality type are bad. As a consequence, this algorithm does not send recommendations for the B and C personality types (which are equivalent). The \textit{Triage} algorithm is the optimal triage algorithm ($\ftriage$) for a rational agent with the subjects' private information.

\item[\textit{Complementary Triage}.] These subjects observe an algorithm that combines the logic of the \textit{Triage} and \textit{Complementary} algorithms. These subjects receive the \textit{Complementary} algorithm's recommendation of hire if at least half of the Communications applicants associated with the given personality type are good and receive the \textit{Complementary} algorithm's recommendation of not hire if at least half of the Sales applicants associated with the given personality type are bad. The \textit{Complementary Triage} algorithm is the optimal algorithmic triage over the subpopulation of Sales and Communications applicants.
\end{description}

We summarize the logic behind these algorithms by examining their implicit assumptions about the subjects' compliance choices and active decisions.
Specifically, \autoref{ms-tab:OptimalAlgorithms} presents each algorithm as a solution to a specific combination of assumptions.
For compliance choices, we distinguish between perfect compliance (that is, following the recommendation and only making an active decision when no recommendation is given) and selective compliance.
By selective compliance, we mean that subjects always make an active decision for Engineering candidates and comply otherwise.
For active decisions, we distinguish between completely random decisions and sophisticated decisions.
Sophisticated decisions assume that the subject uses their private role information effectively when they take active decisions, that is, only hiring Engineering and Sales candidates, which is optimal absent recommendations and across each of our treatment arms.
For example, if compliance is perfect and decisions are sophisticated, then we assume that subjects follow all recommendations and correctly decide in the absence of a recommendation when they do not receive one.

Using these combinations of behaviors, we can interpret the above treatment arms as optimal recommendation algorithms.
First, the \textit{Predictive} algorithm is optimal with perfect compliance and random active decisions.
Second, the assumptions under which the \textit{Triage} algorithm is optimal are that compliance is perfect and active decisions are sophisticated.
In this case, the subject follows the recommendation when it is given and otherwise relies on their private information to decide.
Third, the \textit{Complementary} algorithm is optimal when subjects comply for non-Engineering candidates only, and make random decisions when no recommendation is sent.
Fourth, the \textit{Complementary Triage} recommendations achieve optimal outcomes when subjects do not comply for (and always hire) Engineering candidates, and effectively use their private information to make a sophisticated decision when they do not receive a recommendation. If the human were replaced by a rational agent, this treatment arm would achieve the best performance on average, with only one mistake (not hiring a type-E Communications candidate).

\begin{table}
    \centering
    \begin{tabular}{lcc}
        \toprule
        & \multicolumn{2}{c}{Compliance $C$} \\
        \cmidrule{2-3}
        Active decision $\Dactive$ & Perfect & Selective \\
        \midrule
        Random & \textit{Predictive} & \textit{Complementary} \\
        Sophisticated & \textit{Triage} & \textit{Complementary Triage} \\
        \bottomrule
    \end{tabular}
    \caption{Optimal algorithms under different combinations of compliance choices and active decisions.}
    \label{ms-tab:OptimalAlgorithms}
\end{table}

Although final recommendations only depend on the personality type $X$, we note that constructing these algorithms requires an understanding of the joint distribution of $X$ with the ability $Y^*$ as well as the role $U$. Whether such a joint distribution can be learned in practice depends on the type and quality of data that is available at the time of the construction of the algorithm. Here, we focus instead on an analysis of the relative performance of different recommendation algorithms in a setting where we (and the study subjects) know the joint distribution of $(Y^*,X,U)$. In \autoref{ms-sec:estimating}, we will return to the question of how to design recommendations when the distribution of $U$ remains unknown and our only insight into the humans' behavior is through their unassisted decisions ($\Dnull$).

\subsection{Data and Main Results}

Before analyzing the results of the experiment, we describe the data collection process, final sample sizes for each treatment, and basic performance statistics.
Among 1675 surveys begun, about 35\% failed to answer at least one of the initial comprehension questions and were screened out, leaving us with 1083 subjects that completed our study.
Subjects were split approximately equally between all arms except \textit{Control}, which received half as many observations as the other treatments. Specifically, each of the four treatment arms contained approximately 240 participants who, like the approximately 120 \textit{Control} subjects who did not get recommendations, each took 25 hiring decisions for a total of around 6,000 decision responses per treatment and 3,000 for \textit{Control}.
Overall, the subjects performed well in the experiment and within the estimated time frame.
The subjects of our experiment made the correct decision 78.1\% of the time. The average duration of the experiment was around twelve minutes. However, both the performance (standard deviation of 13 percentage points) and duration (standard deviation of almost eight minutes) show significant heterogeneity across subjects.

We compare treatments by average subject performance, with the main results reported in \autoref{ms-tab:results}.
Subjects in the \textit{Predictive} treatment, who received recommendations from an algorithm that was trained to predict outcomes, got around 76\% of hiring decisions correct, significantly above the around 70\% in the \textit{Control} group that was not supplied with any recommendations.
The \textit{Triage}, \textit{Complementary}, and \textit{Complementary Triage} treatments were all designed to provide recommendations that complement human decisions, and subjects in these groups on average outperformed subjects in the \textit{Predictive} treatment.
Hence, the data confirms that the algorithm designed to take optimal decisions by itself (given by the \textit{Predictive} treatment) is not the optimal recommendation algorithm in this case.
Across the treatments, subjects assigned to the \textit{Complementary} algorithm perform best, with over 81\% optimal decisions (although this performance is statistically indistinguishable from the \textit{Complementary Triage} subjects).
Overall, the results suggest that subjects tend to make smart compliance choices and frequently (and correctly) override the algorithm for Engineering candidates.
At the same time, decisions are inefficiently noisy.

\begin{table}[]
    \centering
    \begin{tabular}{lrlrlrl}
        \toprule
        Treatment arm & Optimal (\%) & (SE) & Hire (\%) & (SE) & Deviated (\%) & (SE) \\
        \midrule
        \textit{Control}  & 69.5 & (0.7) & 67.0 & (1.1) & N/A & \\
        \textit{Predictive}   & 76.2 & (0.5) & 75.3 & (0.7) & 19.7 & (0.9)\\
        \textit{Complementary} & 81.6 & (0.7) & 66.9 & (0.6) & 25.6 & (0.8)\\
        \textit{Triage} & 78.8 & (0.8) &65.9 & (0.6) & 23.5 & (1.0)\\
        \textit{Complementary Triage} & 80.1 & (1.1) & 60.9 & (0.8) & 28.2 & (1.0)\\
        \bottomrule
    \end{tabular}
    \caption{For each treatment we list the percentage of subjects' decisions that were (i) optimal, (ii) hire, or (iii) deviating from the recommendation. Bootstrap standard errors are clustered at the subject level.}
    \label{ms-tab:results}
\end{table}

\begin{figure}
    \centering
    \includegraphics[width = 0.9\textwidth]{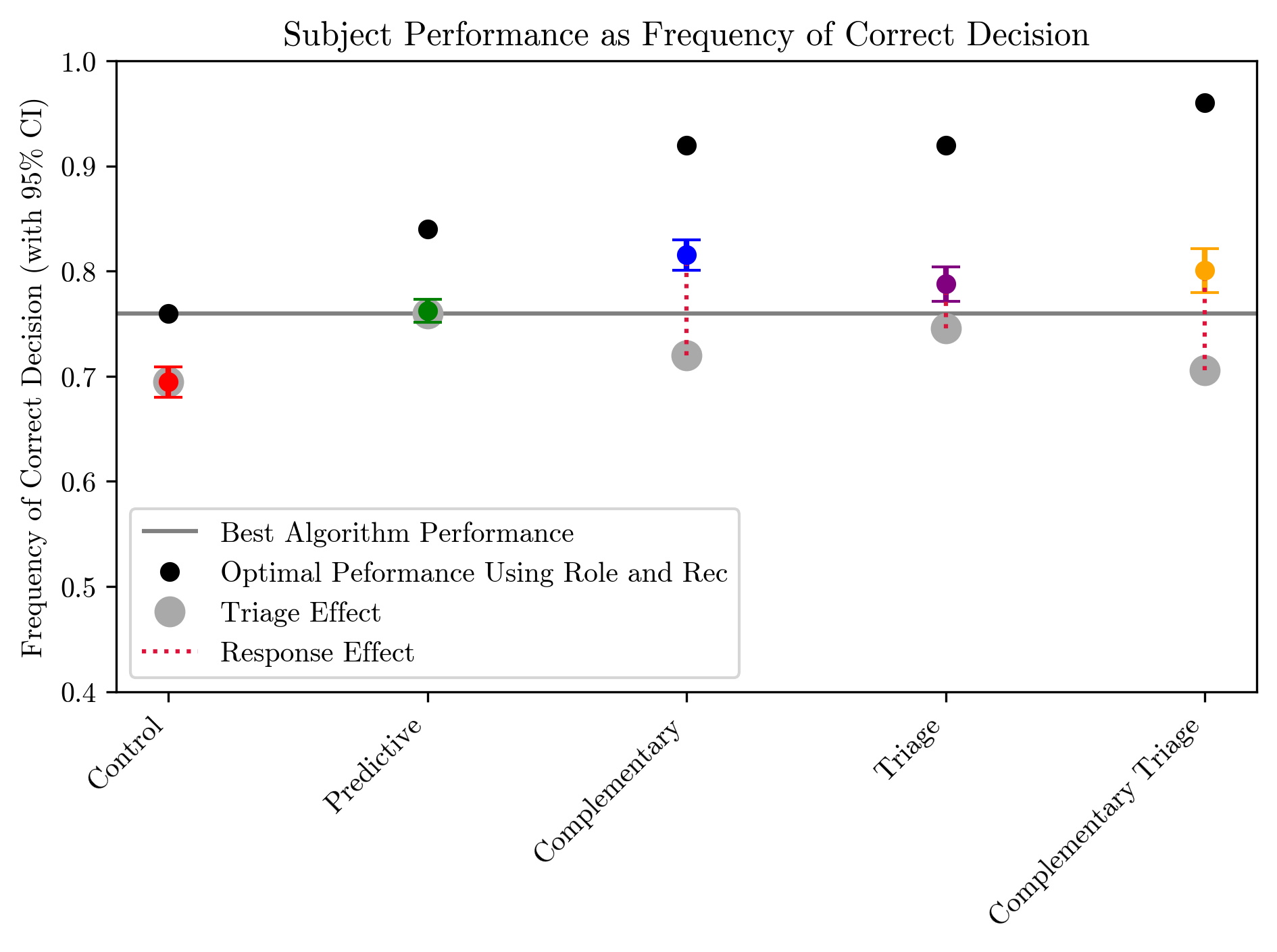}
    \caption{Average subject performance (through fractions of optimal hiring decisions) across treatments. $95\%$ bootstrap confidence intervals are clustered at the subject level. In addition, the triage performance (where compliance is perfect and not sending a recommendation uses the decision of \textit{Control}) is given for each treatment by a gray dot, while the Response Effect from \autoref{ms-subsec:Objectives} is shown by the vertical red dashed lines. The optimal algorithm performance is shown by the gray horizontal line. The optimal performance achievable by a rational agent with the subject's knowledge (role and algorithmic recommendation) is given for each treatment arm as a black dot.}
    \label{ms-fig:results}
\end{figure}

We compare subject performance to useful reference points in \autoref{ms-fig:results} \citep[justified further in][]{wu_rational_2023}. Throughout, we focus on the fraction of cases in which subjects make the optimal decision on an applicant (hire if good, not hire if bad). Within each treatment, we compare subject performance to the performance when following the algorithm perfectly (corresponding to the Triage Effect of \autoref{ms-subsec:Objectives}), and plot the impact of subject deviations (Response Effect). We also provide the best achievable performance by an algorithm acting alone (to benchmark which treatments were able to outperform algorithmic automation) as well as the performance achievable by a rational agent acting with the subject's information (to act as a performance upper bound).
The decomposition reveals that subject performance tends to improve by deviating from the algorithm (at least in cases other than the \textit{Predictive} arm, in which case subject performance is equivalent to perfectly following the algorithm).
Interestingly, machine-assisted human performance is better for the somewhat simpler \textit{Complementary} algorithm than for the theoretically optimal \textit{Complementary Triage}. Overall, we think that the patterns in \autoref{ms-fig:results} are in line with good compliance choices, noisy active decisions, and a penalty for overly complex recommendation algorithms.

Our experiment documents complementarity between humans and algorithms when recommendations are designed well.
Subjects in the \textit{Triage}, \textit{Complementary}, and \textit{Complementary Triage} treatments all performed significantly better (at the 5\% significance level) than humans by themselves (\textit{Control}) or the algorithm by itself, thus achieving human--AI complementarity, while subjects in the \textit{Predictive} treatment could only match the performance humans or algorithms could (theoretically) achieve by themselves.

\subsection{Additional Analysis and Robustness Checks}

\begin{figure}
    \centering
    \includegraphics[width=\textwidth]{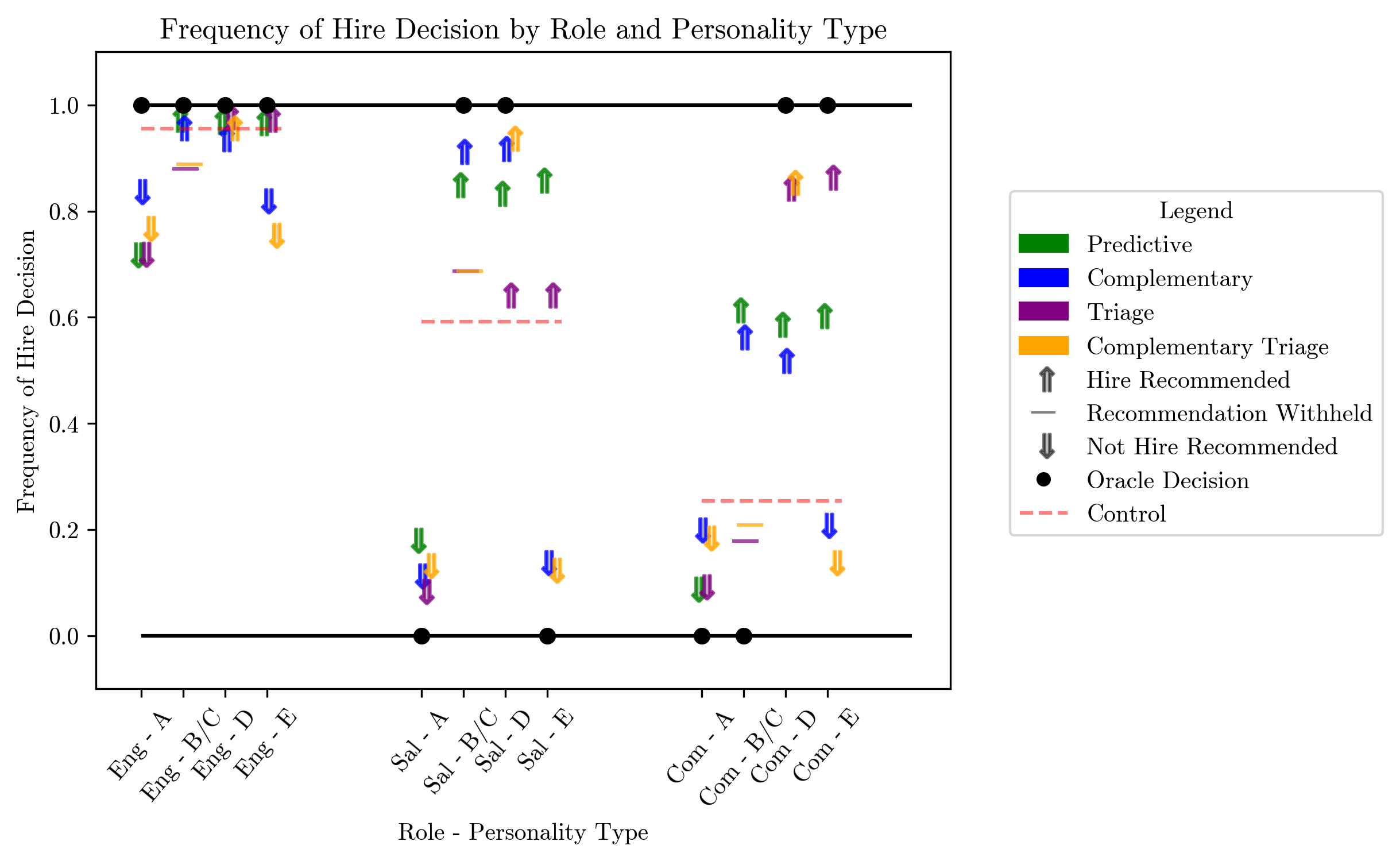}

    \caption{Frequency of experiment subjects taking a hire decision ($y$-axis) across the different combinations of treatments (color) and profiles ($x$-axis). Different markers are used to identify whether the subject received a hire recommendation ($\Uparrow$), did not receive a recommendation ($-$), or received a not-hire recommendation ($\Downarrow$). The optimal decision given oracle information is given by the location of the points on the lines $y=1$ (hire is optimal) and $y=0$ (not hire is optimal). The dashed lines give the frequency of an applicant being hired by a subject without access to recommendations (\textit{Control} arm). }
    \label{ms-fig:heatmap}
\end{figure}

To understand performance differences across treatment arms further, we separate subjects' propensity to hire applicants across roles and personality types in \autoref{ms-fig:heatmap}, and relate this behavior to the recommendations subjects receive.
From the figure, we can identify that subjects' unassisted decisions mostly match the average quality of candidates (which would be in line with a probability-matching model where subjects match the probability they hire an applicant with the probability they are good).
Their compliance choices are generally sophisticated (in that they correctly override the algorithm for Engineering candidates), although somewhat random.
Note that our framework predicts that the \textit{Complementary} algorithm is optimal if subjects make somewhat noisy active decisions but make selective compliance choices.
This is in line with our empirical findings, where the \textit{Complementary}  algorithm is the best among those algorithms we tested and likely the optimal recommendation algorithm we could have implemented.

As an additional robustness check for our results in this section, we exploit the additional comprehension questions we asked subjects about the recommendation algorithm they used. Specifically, 376 of the 961 study subjects who were given a recommendation did not answer one of the comprehension questions correctly about their specific algorithm. We reproduce all analysis on both the sub-population of subjects who answered all recommendation comprehension questions correctly and those who answered at least one question wrong in \autoref{ms-sec:EmpApp}. Our results strengthen on the sub-population who answered all comprehension questions correctly, while, as expected, results are very noisy and largely insignificant on the sub-population of subjects who answered at least one question wrong. These results support conclusions by \cite{bansal_beyond_2019} that a subject's understanding of an algorithm is vital to achieving complementarity.

\section{\texorpdfstring{Designing Recommendations from Data}{Designing Recommendations from Data}}
\label{ms-sec:estimating}

Given our descriptive framework connecting recommendation algorithm design to the human decision-maker's counterfactual decisions, an immediate prescriptive question arises: how should we design recommendation algorithms in practice based on observed data only?
Above, we showed the value of designing recommendation algorithms that achieve complementarity in a lab experiment where we know the joint distribution of the human's information and the algorithm's information.
We now study how our approach can be helpful in the realistic case where we want to design recommendations based on baseline data.
To do so, we first outline challenges that arise when we train recommendation algorithms when only unassisted decision data is available.
We then propose a maximin approach to address these challenges.
Our maximin solution has a particularly intuitive structure: it yields an algorithm that learns where the human decision-maker is likely to make a mistake and then suggests an alternative course of action in those cases.
We then consider an alternative strategy that is motivated by the connection to instrumental variables, an intent-to-treat strategy based on randomized recommendations, link the solutions back to the experiment from the previous section, and discuss machine-learning implementations.
All strategies are instances of one problem. The designer would like to know the joint distribution of $(Y(1), Y(0), X, D_f(\cdot))$, which under \autoref{ms-asm:Mon} is $(Y(1), Y(0), X, \Dactive_f, C_f)$, and chooses the rule that is best in the worst case over the distributions consistent with what is known,
\begin{equation}
\label{ms-eqn:maximin}
    \max_{f \in \F} \ \min_{\P \in \mathcal{P}} \ \E[Y(D_f(R))] \quad \text{for } R = f(X),
\end{equation}
where $\mathcal{P}$ is the set of distributions over $(Y(1), Y(0), X, D_f(\cdot))_{f \in \F}$ that are consistent with the available data and the maintained assumptions. The strategies differ in which assumptions and which data they combine. The assumptions are monotonicity (\autoref{ms-asm:Mon}) and three assumptions introduced below, on the incidence of compliance, on the quality of active decisions, and on invariance across rules. The data conditions are \emph{observed outcomes}, where the joint distribution of $(Y(1), Y(0), X, \Dnull)$ is known, as in the experiment, and a \emph{randomized experiment}, where decisions and outcomes are observed under several randomized rules.

\subsection{Challenges in Learning Recommendation Algorithms from Training Data}
\label{ms-sec:challenges}

We consider the task of learning a good recommendation algorithm from baseline training data in which we observe characteristics $X$, potential outcomes $(Y(1),Y(0))$, as well as unassisted human decisions $\Dnull$, taken in the absence of algorithmic recommendations (corresponding to the \textit{Control} group in the experiment). We abstract from the task of learning that distribution and assume for now that we know the full joint distribution of
\begin{align*}
    (Y(1),Y(0),X,\Dnull).
\end{align*}
This is the observed-outcomes data condition.
Knowing this distribution, how can we design a good recommendation algorithm?

In general, finding an optimal recommendation algorithm from this baseline data alone is not feasible since relevant parts of the potential-decisions distribution are not identified based on this data alone. Just assuming monotonicity (\autoref{ms-asm:Mon}) does not alleviate the problem. Specifically, we face the following four practical challenges:
\begin{enumerate}
    \item \emph{No information about compliance.} Since this baseline data does not include a recommendation algorithm, we cannot learn about compliance. Below, we, therefore, make assumptions about the incidence of compliance that express the idea that the decision-maker is more likely to comply in cases where they would have made a mistake, which we justify by relating mistakes to decision-maker uncertainty.
    \item \emph{Recommendations may affect active decisions.} The introduction of a recommendation algorithm may not just affect decisions for compliers, it may also affect active decisions relative to unassisted decisions. Below, we specifically assume that active decisions may improve over unassisted decisions since \emph{not} giving a recommendation may itself provide additional information to the decision-maker.
    \item \emph{Compliance and active decisions may vary with the recommendation algorithm.} Even if we can make informed guesses about compliance and active decisions for a specific recommendation algorithm, both may change when a different set of recommendations is given. This phenomenon is related to the concept of \emph{interference} in causal inference: the realization of outcomes (here: decisions) for a specific instance may not only depend on the treatment (here: recommendation) for this instance, but also on the treatment status of other instances. In our formalism, we express this challenge by the explicit dependence of potential decisions $D_{f}(\cdot)$ (as well as compliance $C=C(f)$ and active decisions $\Dactive = \Dactive(f)$) on the recommendation algorithm $f \in \F$, such as in \autoref{ms-asm:Mon} and \autoref{ms-prop:simpPO}. Below, we address this challenge by making assumptions that hold across (reasonable) recommendation algorithms $\F$.
    \item \emph{Outcomes are selectively observed.} Data on unassisted decisions reveal $Y(1)$ only where $\Dnull = 1$ and $Y(0)$ only where $\Dnull = 0$, and $\Dnull$ depends on the private information $U$.
\end{enumerate}
These challenges imply that we are not generally able to solve for optimal recommendation algorithms based on baseline data alone. Instead, we have to combine assumptions on compliance, active decisions, and interference.
In the spirit of our overall framework, we aim to make relatively weak assumptions.
This means there is remaining uncertainty, which we resolve by searching for maximin solutions that make worst-case assumptions about unidentified parts of the distribution of potential decisions.

\subsection{\texorpdfstring{Learning from Mistakes with Observed Outcomes}{Learning from Mistakes with Observed Outcomes}}
\label{ms-sec:design-lfm}

We first consider the observed-outcomes case, in which the designer knows the joint distribution of $(Y(1), Y(0), X, \Dnull)$. Define the \emph{mistake} of the unassisted decision as the gain from reversing it,
\[
    M = Y(1 - \Dnull) - Y(\Dnull),
\]
which equals $\tau = Y(1) - Y(0)$ when $\Dnull = 0$ and $-\tau$ when $\Dnull = 1$, and write $m_0(x) = \E[M \mid \Dnull = 0, X = x]$, $m_1(x) = \E[M \mid \Dnull = 1, X = x]$ and $p_d(x) = \P(\Dnull = d \mid X = x)$. In the observed-outcomes case these objects are known. A mistake in this sense is a causal effect: the effect of overturning the human's decision for the instances the human treated in a given way. We impose two assumptions, one on which instances comply and one on active decisions.

\begin{asm}[Incidence of compliance]
\label{ms-asm:incidence}
For all $f \in \F$, $d \in \{0, 1\}$ and $x$, compliance is non-negatively correlated with mistakes and non-trivial: $\Cov(C_f, M \mid \Dnull = d, X = x) \geq 0$ and $\P(C_f = 1 \mid \Dnull = d, X = x) \geq \varepsilon$ for some $\varepsilon > 0$.
\end{asm}

\noindent The assumption expresses the idea that the decision-maker is more likely to change their decision in response to a recommendation in cases where they would have made a mistake without assistance.

\begin{asm}[Active vs baseline decision quality]
\label{ms-asm:active}
    For all recommendation algorithms $f \in \F$,
    \begin{equation*}
        \E\left[
            Y(\Dactive) - Y(\Dnull) \middle| C {=} 0 \text{ or } R {=} \Neutral \right] \geq 0.
    \end{equation*}
\end{asm}

\begin{prop}[Learning from mistakes]
\label{ms-prop:lfm}
Under \autoref{ms-asm:Mon}, \autoref{ms-asm:incidence} and \autoref{ms-asm:active}, and with the joint distribution of $(Y(1), Y(0), X, \Dnull)$ known, the worst-case gain over the unassisted baseline of recommending $r$ at $x$ is $\varepsilon \, \Psi(r, x)$ if $\Psi(r, x) > 0$ and $\Psi(r, x)$ otherwise, where $\Psi(\Neutral, x) = 0$, $\Psi(1, x) = p_0(x) \, m_0(x)$ and $\Psi(0, x) = p_1(x) \, m_1(x)$. The rule
\[
    f^{\textnormal{LFM}}(x) \in \argmax_{r \in \{0, \Neutral, 1\}} \Psi(r, x)
\]
is therefore maximin optimal in \eqref{ms-eqn:maximin}: it recommends treatment where the instances the human leaves untreated would on average gain from treatment, recommends no treatment where the instances the human treats would on average gain from no treatment, and otherwise sends no recommendation.
\end{prop}

\noindent The rule coincides with the triage rule \eqref{ms-eqn:triage} (see also \autoref{ms-prop:trilfm}), which is no coincidence: \autoref{ms-asm:incidence} allows compliance to be uncorrelated with mistakes, and the worst case is attained there.

\paragraph{The binary-outcome case} In the hiring example of \autoref{ms-sec:setup}, $M$ takes the values $c_I = 1 - c$ for a missed good hire, $c_{II} = c$ for a bad hire, and $-c_{II}$ or $-c_I$ when the unassisted decision was right, so that $\E[M \mid \Dnull = 0, X = x] \geq 0$ is $\P(Y^* = 1 \mid \Dnull = 0, X = x) \geq c$. The remainder of this subsection restates the result for this case with the mistake replaced by the indicator $\tilde{M}$ of a mistake and with a parameter $\kappa$ that lets the correlation between compliance and mistakes be weaker or stronger than in \autoref{ms-asm:incidence}.

To determine plausible recommendation algorithms from baseline data about unassisted decisions $\Dnull$, characteristics $X$, and latent outcomes $Y^*$, we now make additional assumptions about how compliance and active decisions relate to unassisted decisions and their performance in the baseline.
We then solve for maximin optimal solutions to resolve additional uncertainty about the full distribution of potential decisions.

Recall that the optimal recommendation algorithm under monotonicity (\autoref{ms-asm:Mon}) maximizes
\begin{equation*}
\E\left[Y\right]
=
    \E\left[Y(R) \: \Ind\left[C {=} 1 \text{ and } R {\neq} \Neutral \right]
    + Y(\Dactive) \: \Ind\left[C {=} 0 \text{ or } R {=} \Neutral\right] \right]
\end{equation*}
under the continued assumptions from \autoref{ms-sec:po}, as in \eqref{ms-eqn:compliancedecomposition}.
We now address these two components separately.
For the second part, we note that we do not directly observe active decisions $\Dactive$, since these may differ from entirely unassisted decisions $\Dnull$ through additional information provided by the recommendation algorithm.
For example, even the (non-)recommendation $R = \Neutral$ may act as a signal that a human decision-maker may leverage together with their private information.
To handle such cases, we assume that active decisions cannot be any worse than unassisted decisions (\autoref{ms-asm:active}).

As a direct consequence, we can bound the gain from implementing an arbitrary recommendation algorithm $f$ relative to the unassisted baseline case from below by
\begin{align*}
     \E\left[Y - Y(\Dnull)\right]
     &\geq
     \E\left[(Y(R) - Y(\Dnull)) \: \Ind\left[C {=} 1 \text{ and } R {\neq} \Neutral \right]\right].
\end{align*}
Maximizing the worst-case gain then corresponds to maximizing the remaining gain stemming from compliers.
To restrict compliance, we consider its relationship to the decision-maker's mistakes in the baseline data.
We define
\begin{align*}
    \tilde{M} &= \Ind[Y(1{-}\Dnull) {>} Y(\Dnull)] = M_I + M_{II},
    &
    M_I &= \Ind[Y^* {=} 1, \Dnull {=} 0],
    &
    M_{II} &= \Ind[Y^* {=} 0, \Dnull {=} 1],
\end{align*}
and call $\tilde{M}$ a \emph{(baseline) mistake}, which covers the mutually exclusive cases of a type-I error ($M_I$) and a type-II error ($M_{II}$).
Were we to assume that, for a specific baseline decision and recommendation, compliance is positively correlated with baseline mistakes,
\begin{equation}
    \label{ms-eqn:complierselection}
    \P(C=1|\tilde{M}=1,\Dnull,R) \geq \P(C=1|\tilde{M}=0,\Dnull,R),
\end{equation}
then the decision-maker would be weakly more likely to change their decision in response to a recommendation in cases where they would have made a mistake without any assistance.
Thus, those cases in which a lot of the same mistakes happen could be improved by suggesting an action that goes in the opposite direction of those mistakes.
Specifically, we will later establish that the recommendation rule
\begin{equation}
    \label{ms-eq:mistakesrule}
        R
        =
        \begin{cases}
            0 & \E[\tilde{M}|\Dnull=1,X] \geq \frac{c_I}{c_{I} + c_{II}} \\
            1 & \E[\tilde{M}|\Dnull=0,X] \geq \frac{c_{II}}{c_{I} + c_{II}}\\
            \Neutral & \text{else}
        \end{cases}
\end{equation}
is maximin optimal (assuming for simplicity that only one of the two conditions is fulfilled at a time).
In words, this rule checks how many mistakes were made in baseline data for a given $X$ and unassisted decision $\Dnull$.
If the probability of a mistake is so high that taking the opposite decisions would improve performance, then \eqref{ms-eqn:complierselection} will ensure that recommending that alternative plan of action cannot make things any worse.

We now express the idea that compliance is related to mistakes in the baseline data in a more flexible way. Specifically, we use an assumption that expresses how strongly compliance correlates with baseline mistakes.
This allows for both expressing assumptions that are weaker than the one in \eqref{ms-eqn:complierselection} as well as assuming that the relationship between compliance and mistakes is stronger.

\begin{asm}[Incidence of compliance]
\label{ms-asm:compliancemistakes}
For all recommendation algorithms $f \in \F$,
    \begin{equation*}
        \P(C=1|\tilde{M}=1,\Dnull,R) \geq \kappa \: \P(C=1|\tilde{M}=0,\Dnull,R).
    \end{equation*}
\end{asm}

This assumption generalizes \eqref{ms-eqn:complierselection}, which is the special case $\kappa = 1$.
The assumption is not restrictive at all for $\kappa = 0$, and for $0 < \kappa < 1$ only assumes that compliance is not creating a large number of new mistakes without also correcting some old mistakes.
For $\kappa > 1$, it assumes that compliance is more likely if a mistake would have been committed in the baseline and is indicative of the decision-maker using their private information to comply selectively.
As $\kappa$ grows large, more and more mistakes are overridden by compliant decisions, while relatively fewer new mistakes are added.
To give this assumption bite in those cases, we also assume that compliance is positive (at least conditional on mistakes):

\begin{asm}[Non-trivial compliance]
\label{ms-asm:complianceminimum}
For all recommendation algorithms $f \in \F$,
    \begin{equation*}
        \P(C=1|\tilde{M}=1,\Dnull,R) \geq \varepsilon.
    \end{equation*}
\end{asm}

Putting all these assumptions together, we can now derive a maximin optimal recommendation algorithm that generalizes the recommendation rule above.

\begin{alg}[Learning from mistakes]
    \label{ms-alg:LFM}
    For a given $\kappa \geq 0$,
    \begin{equation*}
        f^{\textnormal{LFM}}_\kappa(x)
        =
        \begin{cases}
            0 & \text{if only } \E[\tilde{M}|\Dnull=1,X=x] \geq \frac{c_I}{c_{I} + \kappa c_{II}} \\
            1 & \text{if only } \E[\tilde{M}|\Dnull=0,X=x] \geq \frac{c_{II}}{\kappa  c_{I} + c_{II}}\\
            \Neutral & \text{if neither condition holds}
            \\
            \argmax_{r \in \{1,0\}} \Delta(r,x) & \text{if both conditions hold}
        \end{cases}
\end{equation*}
where
\begin{equation*}
    \Delta(r,x)
    =
    \begin{cases}
        \left((c_I+c_{II} \kappa ) \E[\tilde{M}|\Dnull{=}1,X{=}x]  - c_I \right) \P(\Dnull{=}1|X=x), & r = 0 \\
        \left((\kappa c_I+c_{II}) \E[\tilde{M}|\Dnull{=}0,X{=}x]  - c_{II}\right) \P(\Dnull{=}0|X=x), & r = 1.
    \end{cases}
\end{equation*}
\end{alg}

This algorithm learns from mistakes: it sends a recommendation only if the unassisted decisions lead to so many mistakes that giving the opposite recommendation can only improve decisions.
    The structure of the algorithm may look complicated, but is intuitive. The first two rows check whether recommending $0$ or $1$ are guaranteed to improve outcomes relative to unassisted decisions, respectively.
If neither leads to guaranteed improvements in expected outcomes, no recommendation is given; if both lead to improvements, then the one that leads to the larger guaranteed increase is selected.

\begin{thm}
\label{ms-thm:LFM}
    Under \autoref{ms-asm:Mon}, \autoref{ms-asm:active}, \autoref{ms-asm:compliancemistakes} and \autoref{ms-asm:complianceminimum},
    if \autoref{ms-alg:LFM} is in $\F$, then it is maximin optimal in the sense that it solves, for given  $\varepsilon \in (0,1]$ and $\kappa \geq 0$,
    \begin{align*}
        \max_{f \in \F}
        \min_{\P \in \mathcal{P}}
        \E[Y]
        \text{ for } R = f(X), D = D_f(R), Y = Y(D)
    \end{align*}
    where $\mathcal{P}$ are those distributions over $(Y^*,X,\Dnull,D_f(\cdot))$ that fulfill the assumptions and are consistent with the observed distribution of baseline data $(Y^*,X,\Dnull)$.
\end{thm}

We close this part by comparing the resulting algorithm to the alternatives we discuss in \autoref{ms-sec:po}.
Specifically, the automated \emph{decision} from \eqref{ms-eq:OptDec} that an algorithm could take is to hire whenever $\P(Y^*{=}1|X{=}x) \geq \frac{c_{II}}{c_I + c_{II}}$.
While this algorithm ignores human decisions, our proposed approach focuses instead on mistakes that capture the relative performance of human vs machine decisions.
The \emph{triage} rule from \eqref{ms-eqn:triage} decides whether to override human decisions based on whether they perform better or worse than automated algorithmic decisions, and is a special case of \autoref{ms-alg:LFM}.

\begin{restatable}[Triage as learning from mistakes]{prop}{trilfm}
    \label{ms-prop:trilfm}
    If $\kappa = 1$, then $f_\kappa^{LFM} = \ftriage$.
\end{restatable}

The intuition behind this equivalence can be seen by comparing \autoref{ms-alg:LFM} to the following representation of the triage rule,
\begin{equation}
\label{ms-eqn:triage-explicit}
    \ftriage(x)
    =
    \argmax_{r \in \{0,1,\Neutral\}}
    \begin{cases}
        -c_I \P(Y^*{=}1|X{=}x), & r = 0 \\
        -c_{II} \P(Y^*{=}0|X{=}x), & r = 1 \\
        -c_{I} \E[M_I|X{=}x] - c_{II}\E[M_{II}|X{=}x], & r = \Neutral,
    \end{cases}
\end{equation}
where $M_I$ and $M_{II}$ express baseline human type-I and type-II errors as above. Specifically, both formulations show that for a recommendation to be sent, the baseline data must contain a sufficient number of mistakes. If a large number of those mistakes are type-I errors, then both the probability of a mistake given we do not hire and the probability the applicant is good will be high, leading both algorithms to send a hire recommendation. A similar argument holds for type-II errors.

Through \autoref{ms-prop:trilfm}, our approach delivers insights into the connection between recommendations and triage:
\begin{itemize}
    \item If we believe that there is no correlation between compliance and mistakes  ($\kappa = 1$), then optimizing for triage yields a maximin optimal recommendation rule within our framework.

    \item If we believe, however, that compliance is selective in that it is more likely to overturn mistakes than create new ones ($\kappa > 1$), then we obtain a rule that deviates from triage by giving recommendations more often in response to mistakes the unassisted human makes.

    \item If, on the other hand, we are worried that compliance may not be positively correlated with mistakes ($\kappa < 1$), then we may want to withhold more recommendations.

    \end{itemize}
Overall, these distinctions showcase that the central questions in designing algorithms for complementarity are around how the decision-maker leverages their private information to decide when to comply with a recommendation and how their active decision changes in response to changes in the recommendation algorithm.

\subsection{\texorpdfstring{Intent to Treat with Randomized Recommendations}{Intent to Treat with Randomized Recommendations}}
\label{ms-sec:design-itt}

The second strategy replaces assumptions on the incidence of compliance by data on compliance. Suppose the designer has deployed $K$ recommendation rules $f_1, \ldots, f_K \in \F$, one of which may be the null rule $f^{\Neutral}$, assigning each instance to one rule at random, and observes $(Y, D, X)$ together with the rule that was in place. This is the data condition of a randomized experiment, and it is the design of \autoref{ms-sec:experiment}. The strategy needs the response not to depend on which rule is deployed.

\begin{asm}[Invariance across recommendation rules]
\label{ms-asm:invariance}
For all $f, f' \in \F$, $(\Dactive_f, C_f) = (\Dactive_{f'}, C_{f'})$ almost surely.
\end{asm}

\noindent Under \autoref{ms-asm:invariance} we drop the subscript and write $(\Dactive, C)$. Under it, the value of each rule at $x$ is identified by the intent-to-treat comparison across arms, and the designer can choose, for each $x$, the recommendation that performed best.

\begin{prop}[Intent-to-treat strategy]
\label{ms-prop:itt}
Suppose \autoref{ms-asm:Mon} and \autoref{ms-asm:invariance} hold, and let $V_k(x) = \E[Y \mid X = x, \text{rule } f_k]$ denote the average outcome at $x$ under rule $f_k$ in the experiment. Among all rules that at each $x$ send one of the recommendations $f_1(x), \ldots, f_K(x)$, the rule
\[
    f^{\textnormal{ITT}}(x) = f_{k^*(x)}(x), \qquad k^*(x) \in \argmax_{k \in \{1, \ldots, K\}} V_k(x),
\]
maximizes $\E[Y(D_f(R))]$.
\end{prop}

\subsection{Implementation in the Experiment}
\label{ms-sec:application}

Above, we have derived maximin algorithms that lead to a guaranteed improvement in expected outcomes under assumptions on compliance and active decisions. We now apply this family of recommendation algorithms to the experiment in \autoref{ms-sec:experiment}.
This application demonstrates both the features and the limitations of the agnostic maximin approach.

For our application, we assume that we do \emph{not} have any of the information about private information and its relationship to outcomes available, and instead only rely on information about the joint distribution of decisions, characteristics, and outcomes from the baseline (\textit{Control}) arm.
We see this as a conservative reference point that represents the case where we have no or little information about the setting for which we design a recommendation algorithm, making the assumptions from above appropriate.

Per our result above, the main input to finding maximin optimal recommendations using baseline data is the distribution of \emph{mistakes} across characteristics. To calculate the relative improvement of different recommendations, we also require information about unassisted decisions and the distribution of characteristics. These data are represented in \autoref{ms-tab:baselinedata}.
Together, these numbers completely describe the (empirical) baseline distribution.
Here, we take this empirical distribution as given and derive the (sample analog of the) algorithms from \autoref{ms-thm:LFM} for varying values of $\kappa$.

\begin{table}
\centering
\begin{tabular}{l r r r r r}
    \toprule
    & \multicolumn{5}{c}{Personality type ($x$)}\\
    \cmidrule{2-6}
     & A & B & C & D & E  \\
    \midrule
    $\P(X=x)$
    & .20 & .20 & .20 & .20 & .20 \\
    $\P(\Dnull = 1|X=x)$ & 0.67& 0.68& 0.65& 0.68& 0.67 \\
    $\P(Y^*=1|X=x, \Dnull = 0)$ & 0.06 & 0.52 & 0.58 & 1.00 & 0.53 \\
    $\P(Y^*=0|X=x, \Dnull=1)$&0.43  &0.07   & 0.08 &0.00 &  0.36 \\
    \bottomrule
\end{tabular}
\caption{For each personality type $x$, we report the fraction of applicants, the probability that a subject in the \textit{Control} treatment of our experiment hired an applicant, the probability an applicant was good given they were not hired by a subject in the \textit{Control} treatment (type-I error), and the probability an applicant was good given they were hired by a subject in the \textit{Control} treatment (type-II error).}
\label{ms-tab:baselinedata}
\end{table}

The resulting maximin optimal algorithms are represented in \autoref{ms-fig:optimalrecs} for varying values of $\kappa$. Even in the very conservative case of $\kappa \rightarrow 0$, the algorithm recommends hiring applicants of personality type D, since any deviation can only be a mistake.
In this regime, no other recommendations are given.
As $\kappa$ increases, hiring recommendations are also given for types B, C, and E.
Around $\kappa = 1$, this leads to the triage solution from \eqref{ms-eqn:triage-explicit}.
As the minimum correlation between compliance and baseline mistakes increases further, type A is recommended not to be hired (recovering the \emph{Predictive} algorithm), and the recommendation for type E is eventually switched from hiring to not hiring (recovering the \emph{Complementary} algorithm that performs best in the experiment).
This connects to the analysis of different algorithms in the experiment in \autoref{ms-sec:experiment} and especially in \autoref{ms-tab:OptimalAlgorithms}: if we are willing to assume that compliance is selective (which we interpret here as a high $\kappa$), then we expect the \emph{Complementary} algorithm to work well.

\begin{figure}
    \centering

        \begin{tikzpicture}
        \begin{axis}[
            width=10cm, height=8cm,
            xlabel={$\kappa$},
            ylabel={Personality type},
            ytick={1,2,3,4,5},
            yticklabels={A, B, C, D, E},
            ymin=0.5, ymax=5.5,
            xmin=0.1, xmax=10.5,
            xmode=log,
            axis lines=left,
            grid=minor,
            clip=false,
            legend style={at={(1.2,0.8)}, anchor=north west}, %
        ]

        \addplot[dashed,color=gray,forget plot] coordinates {(0.1,1) (10,1)};
            \addplot[dashed,color=gray,forget plot] coordinates {(0.1,2) (10,2)};
            \addplot[dashed,color=gray,forget plot] coordinates {(0.1,3) (10,3)};
            \addplot[dashed,color=gray,forget plot] coordinates {(0.1,4) (10,4)};
            \addplot[dashed,color=gray,forget plot] coordinates {(0.1,5) (10,5)};

            \addplot[line width=1em,color=blue] coordinates {(0.923,2) (10,2)};

        \addplot[line width=1em,color=red] coordinates {(1.326,1) (10,1)};

            \addplot[line width=1em,color=blue] coordinates {(0.724,3) (10,3)};
            \addplot[line width=1em,color=blue] coordinates {(0.1,4) (10,4)};
            \addplot[line width=1em,color=blue] coordinates {(0.886,5) (5.128,5)};

            \addplot[line width=1em,color=red] coordinates {(5.128,5) (10,5)};

            \addlegendentry{hire ($1$)}
         \addlegendentry{not hire ($0$)}

        \end{axis}
    \end{tikzpicture}

    \caption{Recommendations provided by \autoref{ms-alg:LFM} for varying values of $\kappa$ based on the \textit{Control} arm of the experiment. Note that the $x$-axis is on a log scale to highlight the most relevant transitions and because $\kappa$ is a multiplicative factor that can range from 0 to infinity.}
    \label{ms-fig:optimalrecs}
\end{figure}

Note the approach in this section does not have access to any of the information about complementarity that the algorithms in \autoref{ms-sec:experiment} relied on. Learning such additional information and reasoning about compliance and active decisions beyond the arguments above would require access to data about how decisions change with different recommendations, such as those from the treatment arms other than the \emph{Control}.

Overall, this application shows both strengths and limitations of the approach behind \autoref{ms-alg:LFM}.
Even with limited data, our framework can derive sensible recommendation algorithms.
At the same time, the approach is inherently limited since it does not rely on any learning about how decisions change in response to recommendations in the given setting and instead is sensitive to specific assumptions.
As a result, we obtain both algorithms that improve only somewhat over the baseline (like the \textit{Predictive} algorithm we recover for intermediate $\kappa$) or improve strongly (the \textit{Complementary} algorithm for high $\kappa$), but could not have told based on baseline data alone which algorithm would have done better.
Ultimately, finding better recommendation algorithms will involve adaptive experimentation that learns about the structure of compliance and active decisions.
The intent-to-treat strategy of \autoref{ms-sec:design-itt} is the first step of such a program. With the five arms of the experiment as the deployed rules, each personality type received two or three distinct recommendations across arms, and choosing for each type the recommendation of the arm with the highest accuracy on that type gives the rule in \autoref{ms-tab:itt}: not hire for type A, no recommendation for B and C, and hire for D and E, which is the \textit{Triage} algorithm. Its validity rests on \autoref{ms-asm:invariance}.

\begin{table}
    \centering
    \begin{tabular}{ccc}
        \toprule
        Personality type & Recommendations across arms & Intent-to-treat rule \\
        \midrule
        A & $\Neutral$, 0 & 0 \\
        B & 1, $\Neutral$ & $\Neutral$ \\
        C & 1, $\Neutral$ & $\Neutral$ \\
        D & 1, $\Neutral$ & 1 \\
        E & 1, $\Neutral$, 0 & 1 \\
        \bottomrule
    \end{tabular}
    \caption{Intent-to-treat rule from the experiment. For each personality type, the recommendations sent by the five arms (\autoref{ms-tab:treatments}) and the recommendation of the arm with the highest accuracy on that type.}
    \label{ms-tab:itt}
\end{table}

\subsection{Machine-Learning Implementation}

We close this section by discussing the practical implementation of recommendation algorithms using machine learning.
We assumed above knowledge of the joint distribution of unassisted decisions, characteristics, and outcomes.
In practice, these relationships would have to be learned from data.
Fortunately, the algorithm above is instructive about how this learning could be implemented effectively.
Specifically, the main ingredients to constructing \autoref{ms-alg:LFM} are the probability of mistakes conditional on characteristics and unassisted decisions as well as the baseline decisions.
These correspond to the solution of  straightforward classification problems that estimate
\begin{align*}
    m_0(x) &= \E[\tilde{M}|\Dnull {=} 0,X{=}x],
    &
    m_1(x) &= \E[\tilde{M}|\Dnull {=} 1,X{=}x],
    &
    h(x) &= \P(\Dnull = 1|X{=}x).
\end{align*}
Based on estimates $\hat{m}_0, \hat{m}_1, \hat{h}$, a plug-in rule is
\begin{align*}
    \hat{f}(x)
    =
    \argmax_{r \in \{0,1,\Neutral\}}
    \begin{cases}
        ((c_I + \kappa c_{II}) \hat{m}_1(x) - c_{I}) \hat{h}(x), & r = 0 \\
        ((\kappa c_I +  c_{II}) \hat{m}_0(x) - c_{II}) (1-\hat{h}(x)), & r = 1 \\
        0, & r = \Neutral.
    \end{cases}
\end{align*}
An alternative is to estimate the rule directly via cost-weighted classification by solving the sample analog of
\begin{align*}
    \max_{f \in \F}
    \E\left[
        \Ind[f(X) {=} 1]
        ((\kappa c_I + c_{II}) M_I - c_{II} \Ind[\Dnull {=} 0])
        +
        \Ind[f(X) {=} 0]
        ((c_I + \kappa c_{II}) M_{II} - c_I \Ind[\Dnull {=} 1])
    \right]
\end{align*}
where $M_I$ and $M_{II}$ are type-I and type-II errors, respectively.

\section{Conclusion}
\label{ms-sec:conclusion}

This article formalizes the analysis and optimal design of recommendation algorithms as a causal design problem with endogenous response, which employs the potential-outcomes framework from causal inference. We draw similarities between the impact that recommendation algorithms have on human decisions in algorithmic design and the effect that instrumental variables have on individual treatment choices in inference problems. We use this connection to decompose the objective function of a recommendation algorithm into algorithmic performance and human response.
Making additional assumptions on reasonable human responses to recommendations, our approach allows us to express recommendation-assisted decisions in terms of compliance and active decisions.
We can then reason about optimal recommendation algorithms in terms of assumptions on each component individually.
We demonstrate the utility of our framework in a controlled online experiment and a maximin optimization application, both of which reinforce that decision-maker responses should be considered when designing recommendation algorithms. Beyond learning from mistakes, randomized recommendations yield an intent-to-treat strategy, with an explicit list of the assumptions and data it needs. Specifically, we document in the experiment that subjects perform better using recommendations that are designed to provide complementary information, relative to algorithms that are instead designed to automate decisions.

\bibliography{bib}

\newpage
\appendix

\begin{center}
    \Large
    \textsc{Appendix}
\end{center}

\section{Proofs}
\label{ms-sec:ProofApp}

\noindent The proofs of \autoref{ms-thm:LFM} and \autoref{ms-prop:trilfm} are written in the loss notation of \autoref{exm:classificationhiring}, with $\ell(Y^*, D) = c_I \Ind[Y^* {=} 1, D {=} 0] + c_{II} \Ind[Y^* {=} 0, D {=} 1]$, and write $Y$ for the ability $Y^*$. Since $Y(D) = (Y^* - c) D$ with $c_I = 1 - c$ and $c_{II} = c$, we have $\ell(Y^*, D) = c_I Y^* - Y(D)$, so that minimizing expected loss is the same as maximizing the expected outcome, and the mistake indicator $\tilde{M}$ is the one defined in \autoref{ms-sec:design-lfm}.

\begin{proof}[Proof of \autoref{ms-prop:simpPO}]
    To show this result we will build a bijection between the potential outcomes $(D_{f,i}(0),D_{f,i}(\Neutral),D_{f,i}(1))$ which exist under \autoref{ms-asm:Mon} and realizations of $(\Dactive{}_i,C_i)$:
    \begin{align*}
        &(0,0,0) & &\longleftrightarrow & &(0,0) \\
        &(0,0,1) & &\longleftrightarrow & &(0,1) \\
        &(0,1,1) & &\longleftrightarrow & &(1,1) \\
        &(1,1,1) & &\longleftrightarrow & &(1,0)
        \qedhere
    \end{align*}
\end{proof}

\begin{proof}[Proof of \autoref{ms-prop:lfm}]
    Fix $f \in \F$ and write $R = f(X)$. By \autoref{ms-prop:simpPO},
    \[
        Y(D) - Y(\Dnull) = C_f \, (Y(R) - Y(\Dnull)) \, \Ind[R \neq \Neutral] + (Y(\Dactive_f) - Y(\Dnull)) \, \Ind[C_f = 0 \text{ or } R = \Neutral].
    \]
    The expectation of the second term is non-negative by \autoref{ms-asm:active}, and it equals zero when $\Dactive_f = \Dnull$, which is admissible. For the first term, $Y(1) - Y(\Dnull) = (1 - \Dnull) \, M$ and $Y(0) - Y(\Dnull) = \Dnull \, M$, so that
    \begin{align*}
        \E\bigl[C_f \, (Y(R) - Y(\Dnull)) \, \Ind[R \neq \Neutral]\bigr] = \sum_x \P(X = x) \Bigl( &\Ind[f(x) = 1] \, p_0(x) \, \E[C_f M \mid \Dnull = 0, X = x] \\
        &+ \Ind[f(x) = 0] \, p_1(x) \, \E[C_f M \mid \Dnull = 1, X = x] \Bigr).
    \end{align*}
    For each $d$ and $x$, the covariance identity gives
    \[
        \E[C_f M \mid \Dnull = d, X = x] = \Cov(C_f, M \mid \Dnull = d, X = x) + \P(C_f = 1 \mid \Dnull = d, X = x) \, m_d(x),
    \]
    where the covariance is non-negative and the probability lies in $[\varepsilon, 1]$ by \autoref{ms-asm:incidence}. If $m_d(x) > 0$, the right-hand side is at least $\varepsilon \, m_d(x)$, with equality when $C_f$ is independent of $M$ given $(\Dnull = d, X = x)$ and $\P(C_f = 1 \mid \Dnull = d, X = x) = \varepsilon$. If $m_d(x) \leq 0$, it is at least $m_d(x)$, with equality when $C_f = 1$ on $\{\Dnull = d, X = x\}$. Both choices satisfy \autoref{ms-asm:incidence}, and they can be made separately for each $(d, x)$ and each $f$ and combined with $\Dactive_f = \Dnull$; by \autoref{ms-prop:simpPO} every such specification of $(\Dactive_f, C_f)$ is consistent with \autoref{ms-asm:Mon}, and it leaves the known distribution of $(Y(1), Y(0), X, \Dnull)$ unchanged. Writing $G(r, x) = \varepsilon \Psi(r, x)$ if $\Psi(r, x) > 0$ and $G(r, x) = \Psi(r, x)$ otherwise, we obtain
    \[
        \min_{\P \in \mathcal{P}} \E[Y(D_f(R)) - Y(\Dnull)] = \sum_x \P(X = x) \, G(f(x), x),
    \]
    which is the first claim. The maximin problem \eqref{ms-eqn:maximin} therefore reduces to maximizing $G(f(x), x)$ separately for each $x$. Since $G(r, x)$ and $\Psi(r, x)$ have the same sign for every $r$, both vanish at $r = \Neutral$, and $G = \varepsilon \Psi$ wherever $\Psi > 0$, the two functions have the same maximizers in $r$, which proves the second claim.
\end{proof}

\begin{proof}[Proof of \autoref{ms-thm:LFM}]
    The optimal recommendation algorithm minimizes
    \begin{align*}
    &\E\left[\ell(Y,D_{f}(R)))\right] - \E\left[\ell(Y,\Dnull))\right]
    \\
    &=
    \E\left[(\ell(Y,R) {-} \ell(Y,\Dnull)) \: \mathbf{1}\left(C {=} 1 \text{ and } R {\neq} \Neutral \right)
    + (\ell(Y,\Dactive) {-}\ell(Y,\Dnull))  \: \mathbf{1}\left(C {=} 0 \text{ or } R {=} \Neutral\right) \right]
    \\
    &\leq
    \E\left[(\ell(Y,R) - \ell(Y,\Dnull))  \: \mathbf{1}\left(C {=} 1 \text{ and } R {\neq} \Neutral \right)\right]
    \\
    &=
    \E\left[\E\left[(\ell(Y,R) - \ell(Y,\Dnull)) \mathbf{1}\left(C {=} 1\right) \middle| R, \Dnull\right] \mathbf{1}\left(R {\notin} \{\Neutral,\Dnull \}\right)\right]
    \end{align*}
    where, for $r \neq d$ and $\bar{p}(r,d) = \P(C {=} 1|\tilde{M}{=}0, R{=}r, \Dnull{=}d)$,
    \begin{align*}
        &\E\left[(\ell(Y,R) - \ell(Y,\Dnull)) \mathbf{1}\left(C {=} 1\right) \middle| R{=}r, \Dnull{=}d\right]
        \\
        &=
        \begin{cases}
            \overbrace{(- c_{II})}^{\mathrlap{\text{gain from mistake corrected}}}
            \P(C {=} 1|\tilde{M}{=}1, R{=}r, \Dnull{=}d)
            \P(\tilde{M}{=}1|R{=}r, \Dnull{=}d)
            \\
            +
            c_{I}
            \P(C {=} 1|\tilde{M}{=}0, R{=}r, \Dnull{=}d)
            \P(\tilde{M}{=}0|R{=}r, \Dnull{=}d),
            &
            d = 1, r = 0 \\
            (- c_{I})
            \P(C {=} 1|\tilde{M}{=}1, R{=}r, \Dnull{=}d)
            \P(\tilde{M}{=}1|R{=}r, \Dnull{=}d)
            \\
            +
            \underbrace{c_{II}}_{\mathrlap{\text{loss from new mistake}}}
            \P(C {=} 1|\tilde{M}{=}0, R{=}r, \Dnull{=}d)
            \P(\tilde{M}{=}0|R{=}r, \Dnull{=}d),
            &
            d = 0, r = 1 \\
        \end{cases}
        \\
        &\leq
        \bar{p}(r,d)
        \begin{cases}
            - c_{II} \kappa \E[\tilde{M}|R{=}r,\Dnull{=}d] + c_{I} (1{-}\E[\tilde{M}|R{=}r,\Dnull{=}d])
            &
            d = 1, r = 0 \\
            - c_{I} \kappa \E[\tilde{M}|R{=}r,\Dnull{=}d] + c_{II} (1{-}\E[\tilde{M}|R{=}r,\Dnull{=}d])
            &
            d = 0, r = 1
        \end{cases}
        \\
        &=
        \bar{p}(r,d)
        \begin{cases}
            c_{I} - (c_I+ \kappa c_{II}) \E[\tilde{M}|R{=}r,\Dnull{=}d]
            &
            d = 1, r = 0 \\
            c_{II} - (\kappa c_I+  
 c_{II}) \E[\tilde{M}|R{=}r,\Dnull{=}d]
            &
            d = 0, r = 1
        \end{cases}
    \end{align*}
    where the lower bound can be attained by assuming that the compliance bound from \autoref{ms-asm:compliancemistakes} is binding.
    Writing $m_d(x) = \E[\tilde{M}|X{=}x,\Dnull{=}d]$,
    we therefore obtain the tight upper bound
    \begin{align*}
        &\E\left[\ell(Y,D_{f}(R)))\right] - \E\left[\ell(Y,\Dnull))\right]
        \\
        &\leq
        \E\big[\bar{p}(0,1) \mathbf{1}(\Dnull{=}1,R{=}0) (c_{I} - ( c_I+ \kappa c_{II}) m_1(X))
        \\
        &
        \phantom{\leq
        \E\big[}
        +
        \bar{p}(1,0)\mathbf{1}(\Dnull{=}0,R{=}1) (c_{II} - (\kappa c_I+ c_{II}) m_0(X))
        \big]
        \\
        &=
        \E\big[\bar{p}(0,1) \mathbf{1}(f(X){=}0) (c_{I} - (c_I+ \kappa c_{II}) m_1(X)) \P(\Dnull{=}1|X)
        \\
        &\phantom{\leq
        \E\big[}
        +
        \bar{p}(1,0)\mathbf{1}(f(X){=}1) (c_{II} - (\kappa c_I+ c_{II}) m_0(X)) \P(\Dnull{=}0|X)
        \big].
    \end{align*}
    To minimize this upper bound, we would want to assign only those instances $x$ to the $0$ recommendation for which $c_{I} - ( c_I+ \kappa c_{II}) m_1(X)$ is negative and only those instances $x$ to the $1$ recommendation for which $c_{II} - (\kappa c_I+  c_{II}) m_0(X)$ is negative (since we can otherwise assign to $\Neutral$).
    As a result, all components are non-positive, and the bound from \autoref{ms-asm:complianceminimum} is binding in the worst case, yielding the (still tight) upper bound (for the non-trivial case of $\kappa > 0$)
    \begin{align*}
        &\E\left[\ell(Y,D_{f}(R)))\right] - \E\left[\ell(Y,\Dnull))\right]
        \\
        &\leq
        \varepsilon / \kappa
        \E\big[\mathbf{1}(f(X){=}0) (c_{I} - (c_I+ \kappa c_{II}) m_1(X)) \P(\Dnull{=}1|X)
        \\
        &\phantom{\leq
        \E\big[}
        +
        \mathbf{1}(f(X){=}1) (c_{II} - (\kappa c_I+  c_{II}) m_0(X)) \P(\Dnull{=}0|X)
        \big].
    \end{align*}
    We now write $h(x) = \P(\Dnull{=}1|X{=}x)$
    For instances $x$ for which both $f(x) \in \{1,0\}$ yield a reduction in loss, an optimal choice is the maximizer of the gain
    \begin{align*}
        \Delta(r,x)
        =
        \begin{cases}
            (c_I+ \kappa c_{II}) m_1(X) h(x) - c_{I} h(x), & r = 0 \\
            (\kappa c_I+ c_{II}) m_0(X) (1-h(x)) - c_{II} (1-h(x)), & r = 1,
        \end{cases}
    \end{align*}
    which together yields \autoref{ms-alg:LFM}.
\end{proof}

\begin{proof}[Proof of \autoref{ms-prop:trilfm}]
    To show this result we will show \autoref{ms-eqn:triage-explicit} recovers \autoref{ms-alg:LFM} for $\kappa = 1$. Do to so we will recover the first three cases of \autoref{ms-alg:LFM} by comparing the terms in \autoref{ms-eqn:triage-explicit} for $r=0$ and $r=1$ to $r = \Neutral$, then recover the last condition by comparing $r=0$ to $r=1$.

    Here we rewrite the condition for $r=0$ to be preferred to $r = \Neutral$ as,
    \begin{align*}
        c_I \P(Y = 1|X=x) &\leq c_I \E\left[M_I|X=x\right] + c_{II}\E\left[M_{II}|X=x\right] \\
        c_I \P(Y = 1|X=x) &\leq c_I \P(Y=1,\Dnull = 0|X=x) + c_{II}\P(Y=0,\Dnull = 1|X=x) \\
        c_I \P(Y = 1,\Dnull = 1|X=x)  &\leq c_{II} \left[\P(\Dnull = 1|X=x) - \P(Y = 1,\Dnull = 1|X=x)\right] \\
        \P(Y = 1|\Dnull = 1,X=x) & \leq \frac{c_{II}}{c_I + c_{II}} \\
        1-\P(Y = 0|\Dnull = 1,X=x) & \leq \frac{c_{II}}{c_I + c_{II}} \\
        \P(Y = 0|\Dnull = 1,X=x) & \geq \frac{c_{I}}{c_I + c_{II}} \\
        \E\left[\tilde{M}|\Dnull = 1,X=x\right]&  \geq \frac{c_{I}}{c_I + c_{II}},
    \end{align*}
    recovering the first case of \autoref{ms-alg:LFM} for $\kappa = 1$. Similarly we can rewrite the condition for $r=1$ to be preferred to $r = \Neutral$ as,
    \begin{align*}
        c_{II} \P(Y = 0|X=x) &\leq c_I \E\left[M_I|X=x\right] + c_{II}\E\left[M_{II}|X=x\right] \\
        c_{II} \P(Y = 0|X=x) &\leq c_I \P(Y=1,\Dnull = 0|X=x) + c_{II}\P(Y=0,\Dnull = 1|X=x) \\
        c_{II} \P(Y = 0,\Dnull = 0|X=x) & \leq c_I \left[ \P(\Dnull = 0|X=x) - \P(Y = 0,\Dnull = 0|X=x)\right] \\
        \P(Y = 0|\Dnull = 0,X=x) &\leq \frac{c_I}{c_I + c_{II}} \\
        1-\P(Y = 1|\Dnull = 0,X=x) &\leq \frac{c_I}{c_I + c_{II}} \\
        \P(Y = 1|\Dnull = 0,X=x) &\geq \frac{c_{II}}{c_I + c_{II}} \\
        \E\left[\tilde{M}|\Dnull = 0,X=x\right]  &\geq \frac{c_{II}}{c_I + c_{II}},
    \end{align*}
    which recovers the second and third cases of \autoref{ms-alg:LFM} for $\kappa = 1$. Now we must consider when both $r = 0$ and $r = 1$ are preferred to $r = \Neutral$ and recover the form of $\Delta(r,x)$ when $\kappa = 1$. To do so we rewrite,
    \begin{align*}
        c_I \P(Y=1|X=x) &= c_I \left[\P(Y=1,\Dnull = 0|X=x) + \P(Y=1,\Dnull = 1|X=x) \right] \\
        &= c_I \left[\E\left[M_I|X=x\right] + \P(\Dnull = 1|X=x) - \E\left[M_{II}|X=x\right] \right]
    \end{align*}
    and
    \begin{align*}
        c_{II} \P(Y=0|X=x) &= c_{II} \left[\P(Y=0,\Dnull = 1|X=x) + \P(Y=0,\Dnull = 0|X=x) \right] \\
        &= c_{II} \left[\E\left[M_{II}|X=x\right] + \P(\Dnull = 0|X=x) - \E\left[M_{I}|X=x\right] \right]
    \end{align*}
    and subtract $c_I \E[M_I|X=x] + c_{II} \E[M_{II}|X=x]$ (the value of $r=\Neutral$), from each to recover and equivalent representation of $\Delta(r,x)$.
\end{proof}

\begin{proof}[Proof of \autoref{ms-prop:itt}]
    Under \autoref{ms-asm:invariance}, write $(\Dactive, C)$ for the common value of $(\Dactive_f, C_f)$ over $f \in \F$. By \autoref{ms-prop:simpPO}, for every rule $g$ the realized outcome is $Y(D_g(g(X))) = h(g(X))$, where
    \[
        h(r) = Y(r) \, \Ind[r \neq \Neutral, \, C = 1] + Y(\Dactive) \, \Ind[r = \Neutral \text{ or } C = 0]
    \]
    is a function of $r$ and of $(Y(1), Y(0), \Dactive, C)$ alone, and does not depend on the rule. Hence, with $W(r, x) = \E[h(r) \mid X = x]$,
    \[
        \E[Y(D_g(g(X)))] = \E\bigl[W(g(X), X)\bigr],
    \]
    and a rule that at each $x$ chooses $g(x)$ from the set $\{f_1(x), \ldots, f_K(x)\}$ to maximize $W(g(x), x)$ maximizes the objective among all rules with this choice set. It remains to identify $W(f_k(x), x)$. In the experiment, the rule $f_k$ is assigned independently of the instance, so the conditional distribution of $(Y(1), Y(0), \Dactive, C)$ given $X = x$ is the same in every arm, and the realized outcome in arm $k$ is $h(f_k(X))$ by \autoref{ms-asm:exclusion}. Therefore $V_k(x) = \E[Y \mid X = x, \text{rule } f_k] = \E[h(f_k(x)) \mid X = x] = W(f_k(x), x)$, and $k^*(x) \in \argmax_k V_k(x)$ selects a recommendation $f_{k^*(x)}(x)$ that maximizes $W(\cdot, x)$ over the deployed recommendations at $x$. The proof uses \autoref{ms-asm:Mon} only through the representation of \autoref{ms-prop:simpPO}; invariance of the potential decisions $D_f(\cdot)$ across rules would suffice in its place.
\end{proof}

\section{Robustness Checks}
\label{ms-sec:EmpApp}

Here we reproduce the results of \autoref{ms-fig:results} and \autoref{ms-fig:heatmap} for the sub-populations who answered all of their recommendation comprehension questions correctly (585 subjects) and those who answered at least one question wrong (376 subjects). Subjects who answered at least one recommendation comprehension question incorrectly exhibited noisier decisions on average than those who answered all questions correctly. We find that algorithms generated using our framework exhibited higher levels of complementarity for the sub-population who correctly answered all comprehension questions while the sub-population who answered at least one comprehension question wrong failed to realize complementarity.

\begin{figure}
    \centering
    \begin{subfigure}[b]{0.65\textwidth}
        \includegraphics[width = \textwidth]{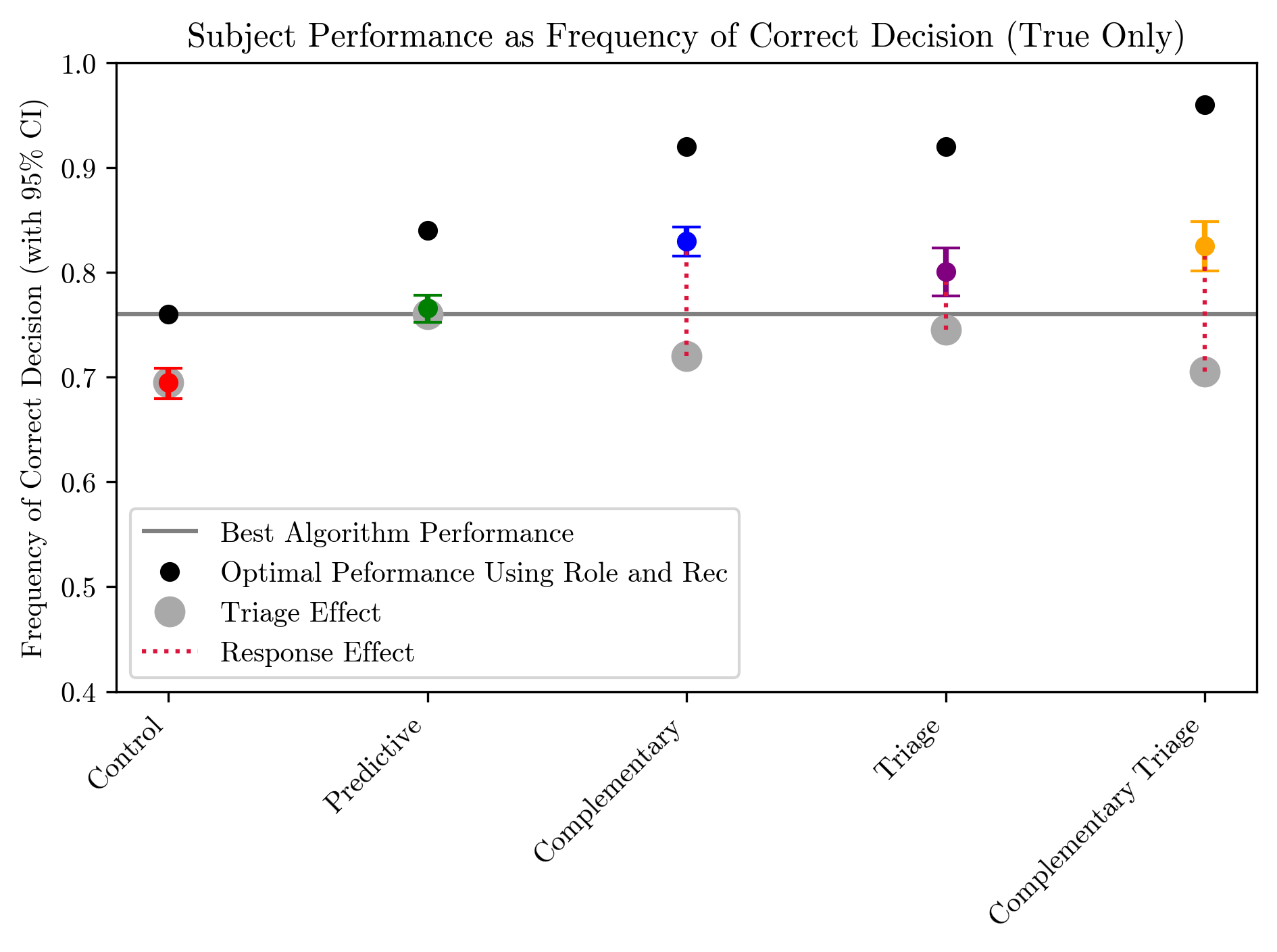}
        \caption{Sub-population with complete comprehension}
        \label{ms-fig:Split4T}
    \end{subfigure}
    \bigskip
    \begin{subfigure}[b]{0.65\textwidth}
        \includegraphics[width = \textwidth]{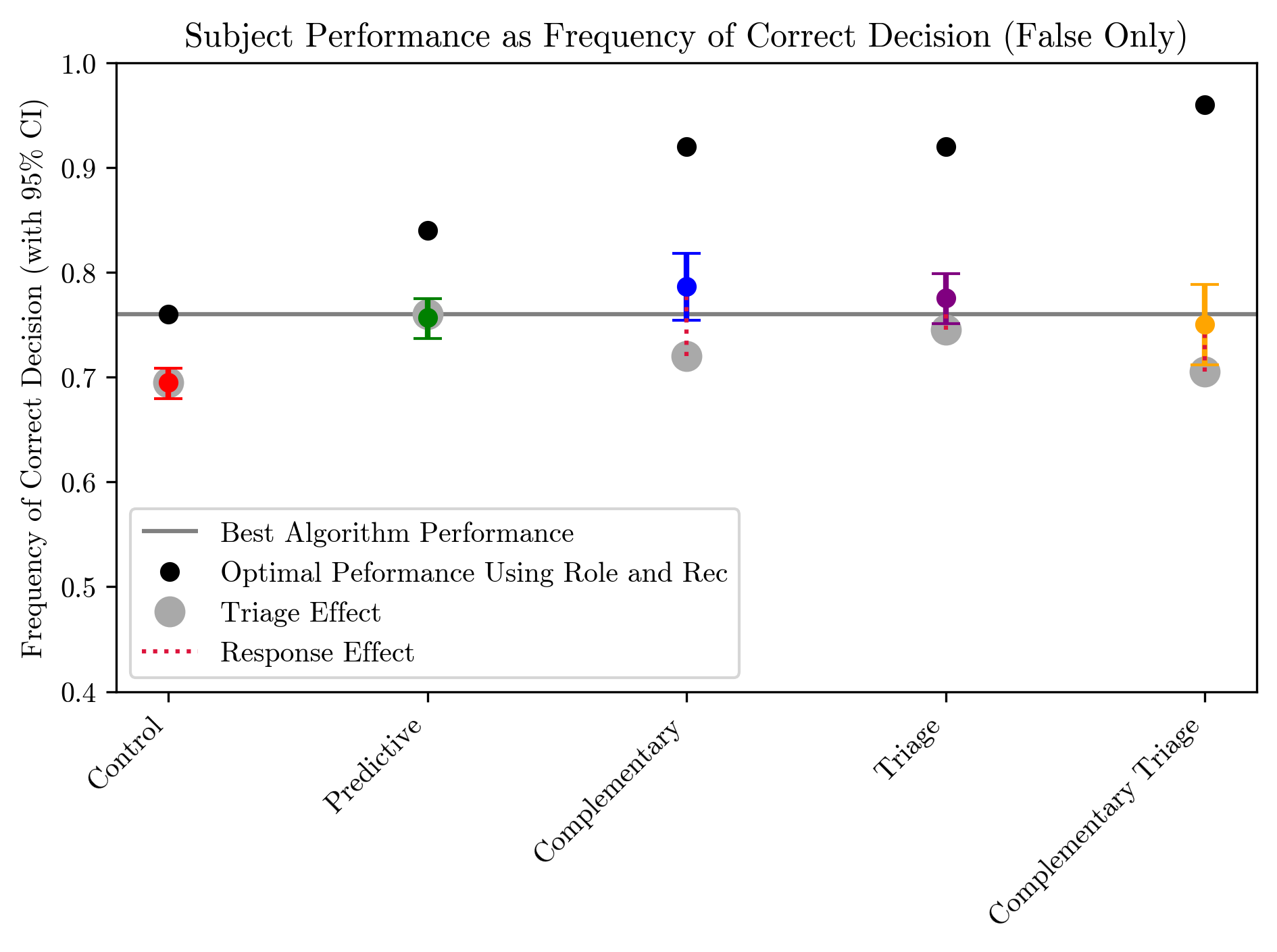}
        \caption{Sub-population with comprehension errors}
        \label{ms-fig:Split4F}
    \end{subfigure}
    \caption{Average subject performance (through fractions of optimal hiring decisions) across treatments. $95\%$ bootstrap confidence intervals are clustered at the subject level. In addition, the triage performance (where compliance is perfect and not sending a recommendation uses the decision of \textit{Control}) is given for each treatment by a gray dot, while the Response Effect from \autoref{ms-subsec:Objectives} is shown by the vertical red dashed lines. The optimal algorithm performance is shown by the gray horizontal line. The optimal performance achievable by a rational agent with the subject's knowledge (role and algorithmic recommendation) is given for each treatment arm as a black dot.}
    \label{ms-fig:Split4}
\end{figure}

\autoref{ms-fig:Split4} reports each sub-populations average performance by treatment. Both sub-populations performed equally well using the \textit{Predictive} algorithm; however, only the sub-population which answered all of our recommendation comprehension questions correctly was able to further improve their performance using the \textit{Triage}, \textit{Complementary}, and \textit{Complementary Triage} algorithms. Performing well with these algorithms required subjects to effectively parse when the recommendation was accurate or erroneous given their private information, thus we expect (and observe) that understanding how the algorithm generates recommendations is vital for its effective usage.

\begin{figure}
    \centering
    \begin{subfigure}[b]{0.75\textwidth}
        \includegraphics[width = \textwidth]{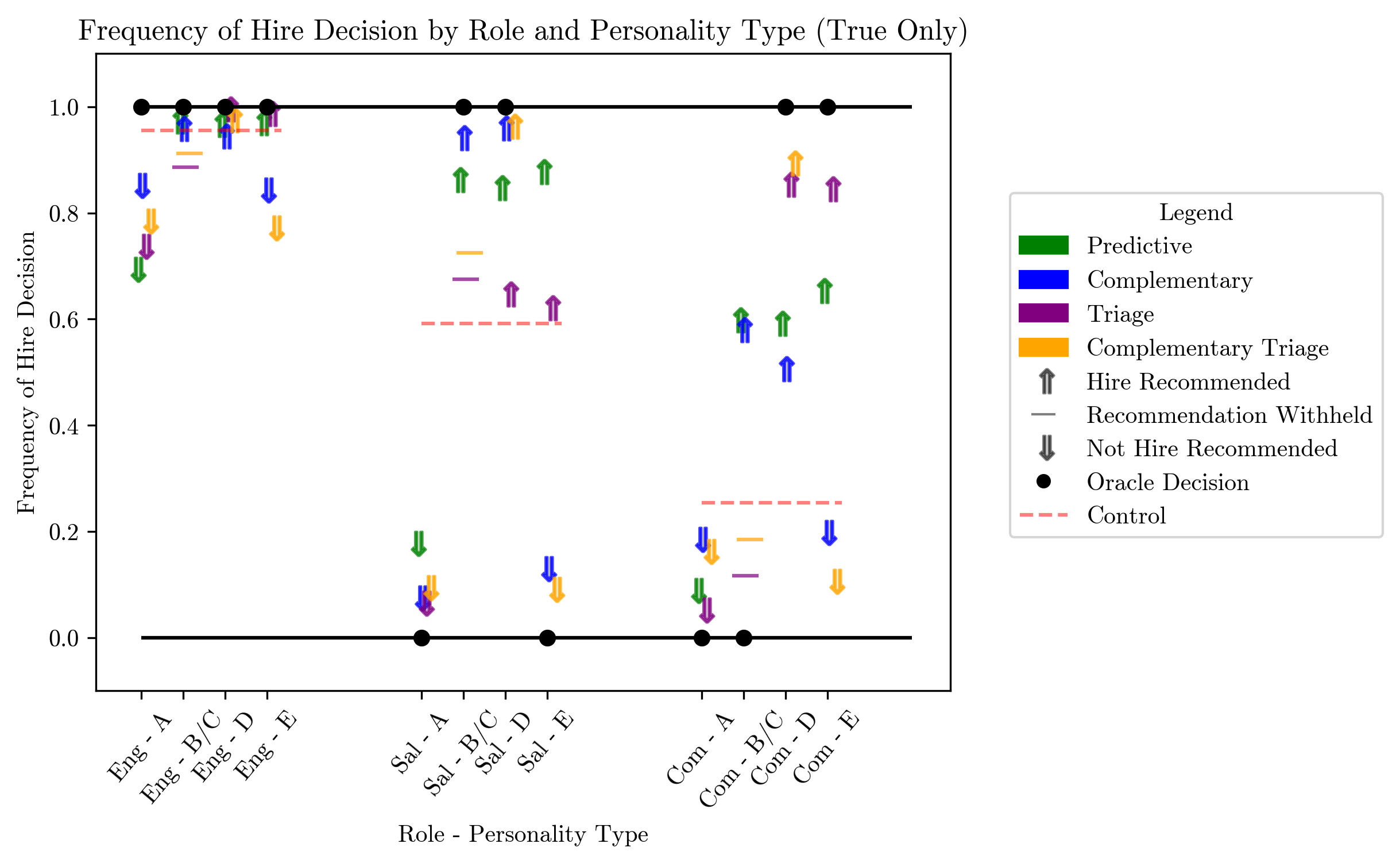}
        \caption{Sub-population with complete comprehension}
        \label{ms-fig:Split5T}
    \end{subfigure}
    \bigskip
    \begin{subfigure}[b]{0.75\textwidth}
        \includegraphics[width = \textwidth]{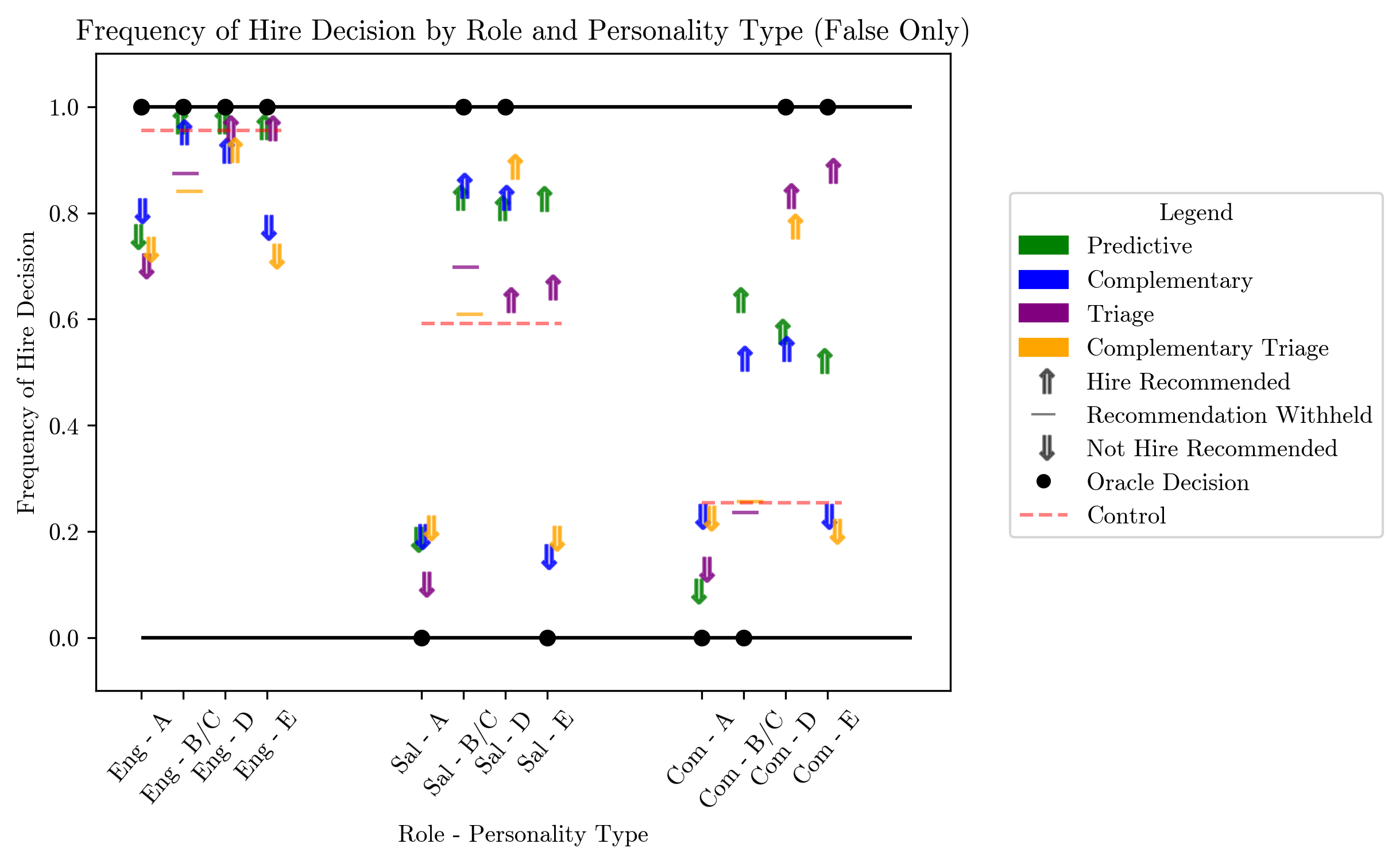}
        \caption{Sub-population with comprehension errors}
        \label{ms-fig:Split5F}
    \end{subfigure}
    \caption{Frequency of experiment subjects taking a hire decision ($y$-axis) across the different combinations of treatments (color) and profiles ($x$-axis). Different markers are used to identify whether the subject received a hire recommendation ($\Uparrow$), did not receive a recommendation ($-$), or received a not-hire recommendation ($\Downarrow$). The optimal decision given oracle information is given by the location of the points on the lines $y=1$ (hire is optimal) and $y=0$ (not hire is optimal). The dashed lines give the frequency of an applicant being hired by a subject without access to recommendations (\textit{Control} arm). The legend for these graphs is the same as the one in \autoref{ms-fig:heatmap} in the main text.}
    \label{ms-fig:Split5}
\end{figure}

\autoref{ms-fig:Split5} reports how each sub-populations propensity to hire applications varied across the profiles they viewed. Generally subjects who answered at least one recommendation comprehension question wrong exhibited noisier decisions. This difference is particularly noticeable on Sales profiles for subjects in the \textit{Complementary} and \textit{Complementary Triage} treatments. Other than this excess noise, subjects who answered at least one recommendation comprehension question wrong responded to recommendations in a similar way to subjects who answered all recommendation comprehension questions correctly.

\end{document}